\documentclass[sigconf]{acmart}

\usepackage{booktabs} 

\setcopyright{none}

\usepackage{amssymb}
\usepackage{graphicx}

\usepackage{url}

\usepackage{multicol}

\usepackage{listings}
\usepackage{xcolor}
\usepackage{xspace}
\usepackage[linesnumbered]{algorithm2e}
\usepackage{caption}
\usepackage{amsfonts}
\usepackage{amsmath}
\usepackage{url}
\usepackage{upgreek}
\usepackage{enumerate}
\usepackage{amssymb}

\newcommand{\code}[1]{\texttt{\footnotesize #1}}

\newcommand{\figref}[1]{Figure~\ref{#1}}
\newcommand{\algref}[1]{Algorithm~\ref{#1}} 
\newcommand{\tblref}[1]{Table~\ref{#1}}

\newcommand{\wdel}[1]{}
\newcommand{\wadd}[1]{#1}

\def\revise#1#2{{\color{red}{\small #1}}
	{{\color{red}\mbox{$\Rightarrow$}}} {\color{blue}{#2}}}
\renewcommand{\revise}[2]{#2}

\begin{document}
\title{Faster Mutation Analysis via\\ Equivalence Modulo States}

%
%
%


\author{Bo Wang, Yingfei Xiong, Yangqingwei Shi, Lu Zhang, Dan Hao\\
	Key Laboratory of High Confidence Software Technologies (Peking University), MoE\\
	Institute of Software, EECS, Peking University, Beijing, 100871, China\\
	\{wangbo\_15, xiongyf, shiyqw, zhanglucs, haodan@pku.edu.cn\}@pku.edu.cn\\
}
\renewcommand{\shortauthors}{Bo Wang et al.}

\begin{abstract}

Mutation analysis has many applications, such as asserting the
quality of test suites and localizing faults. One important
bottleneck of mutation analysis is scalability\wdel{, and researchers have
proposed different techniques to accelerate mutation analysis}. The latest work explores the possibility of reducing the redundant
execution via split-stream execution.\wdel{ That is, all mutates are
	executed as one process and a new process is split from the main
	process when the first mutated statement of a mutant is encountered.}
However, split-stream execution is only able to remove redundant
execution before \wdel{executing }the first mutated statement.

In this paper we try to also reduce some of the redundant execution
after the execution of the first mutated statement. We observe that,
although many mutated statements are not equivalent, 
the execution result of those mutated statements may
still be equivalent to the result of the original statement. In
other words, the statements are equivalent modulo the current state.
\wdel{Based on this observation,} In this paper we propose a fast mutation
analysis approach, AccMut. AccMut automatically detects the
equivalence modulo states among a statement and its mutations, then
groups the statements into equivalence classes modulo states, and
uses only one process to represent each class. In this way, we can
significantly reduce the number of split processes. Our experiments
\wdel{on 5 real-world open source programs with 21558 lines of code, 119176 mutants and 11397 
	tests }show that our approach can further accelerate mutation
analysis on top of split-stream execution\wdel{, a state-of-the-art approach,}
with a speedup of 2.56x on average.
\end{abstract}

%
%
\wdel{

}


\maketitle

\section{Introduction}

\emph{Mutation analysis}~\cite{demillo1978hints,hamlet1977testing,jia2011analysis} is a powerful approach for program analysis. The general process of mutation analysis has two steps. First, change the  original program with predefined mutation operators and generates a set of mutated program, called \emph{mutants}. Then, the mutants are executed against a test suite, and information is collected during the execution for various purpose of analysis.

Mutation analysis has many applications.
The most representative
application is to assess the quality of a test suite. In this
application, the mutants are treated as seeded faults, and the test
suite that detects more mutants is considered better~\cite{Andrews:05,just2014mutants}. A test case
fails on a mutant is known to \emph{kill} that mutant. There are also
many other applications of mutation analysis. For example, recently
several papers \cite{papadakis2012using,moon2014ask,zhang2013injecting,hong2015mutation,icse17FaultPrediction,chekam2016assessing,papadakis2015metallaxis,papadakis2014effective} proposed to
use mutation analysis for fault localization. Bug fixing techniques in
the ``generate-and-validate'' style~\cite{GenProg,PAR,RSRepair} have
been shown to be a dual of mutation analysis~\cite{AE}.
Mutation analysis is also used for test generation~\cite{fraser2012mutation,fraser2015achieving,souza2016strong}
and program verification~\cite{galeotti2015inferring}.
In addition,
variability management and analysis techniques are shown to be quite
close to mutation analysis~\cite{Devroey:2014:VPM:2635868.2666610,devroey2016featured,martin2007fault,papadakis2014sampling}.

However, mutation analysis suffers from one bottleneck: scalability. 
Since we need to test all mutants for a test suite, the expected
analysis time is $n$ times of the expected execution time of the test
suite, where $n$ is the number of mutant generated. 
The number of $n$ depends on the size of the program, but even the mid-size program produces thousands of mutants.
\wdel{ As a result, although mutation analysis has been proposed nearly 40 years, it is adopted only in a limited scope in practice. }

Researchers have realized the problem of scalability, and have
proposed many different approaches for accelerating mutation analysis.
One of the basic ideas of acceleration is to remove the redundant and
unnecessary computations. Mutant schemata~\cite{untch1993mutation} avoids the redundancies in 
compilations. Just et al.'s approach~\cite{just2014efficient} removes redundant mutation
executions while Zhang et al.'s approach~\cite{zhang2013faster} removes unnecessary mutation
executions. The latest work explores split-stream execution~\cite{tokumoto2016muvm} to remove
redundancies in part of the mutation executions. Given two mutants,
the execution of any test before the first mutated statement is
redundant. Split-stream execution starts with one process representing
the original program, and split a new process when the first mutated
statement of a mutant is
encountered. In this way, the redundancies before the first mutated
statement are removed. 

%
%

Split-stream execution only removes redundant executions before the
first mutated statement. However, executions after the first mutated
statement may also be redundant. A typical case is that two statements
are equivalent modulo the current state. Given the current state of
the program, if the executions of two statements lead to the same new
state, we say the two statements are equivalent modulo the state. We
observe that, although in general only a small portion of
mutated statements are equivalent, there are many more mutated
statements that are equivalent modulo the current state. For example,
given two different side-effect-free Boolean expressions, the
probability that they are equivalent is small. However, given a state,
the probability that the two expressions produce the same result is
50\%, if we assume the evaluation result has an even distribution.
Given two mutants and the state before their first different
statements, if the two statements are equivalent modulo the state, the
execution of the two statements and all following execution until the
next different statements are still
redundant. 

In this paper we propose a novel mutation analysis approach, AccMut,
to remove such redundancies. Like split-stream execution, AccMut
starts the executions of all mutants in one process. When we encounter
a position where at least two mutants have different statements,
AccMut clusters the mutants based on their equivalence classes modulo
the current state, i.e., two mutants are put into the same cluster if
their next state after executing the current statement is still the
same. Next, AccMut splits the execution into a set of processes, where
each process represents one cluster of equivalent mutants.

In this
way, AccMut can remove part of the redundancies after the first mutated
statement. More concretely, there are two more types of reductions compared
to split-stream execution. First, when a mutated statement is
equivalent to the original statement modulo the current state, we do
not unnecessarily split the execution from the main process. Second, when two mutated
statements are equivalent modulo the current state, we do not
unnecessarily split two processes for them.
 
	


There are some challenges to implement AccMut.
First, the cluster process, which is invoked at every location
mutation occurs, should be efficient, otherwise the
overhead from the clustering may offset the benefit. In our approach
we have carefully designed the clustering process so that the
time complexity is effectively constant. Second, the process splitting
should also be efficient. In our approach we exploit the POSIX system
call \code{fork} to implement the clustering process. This design
choice has two benefits. First, it allows us to still compile and
execute the mutant, in contrast to existing implementation of
split-stream execution that relies on interpreters. As studied by
existing work~\cite{jia2011analysis,King:1991:FLS:116633.116640}, compiler-based mutation analysis is usually much
faster than interpreter-based. Second, based on the copy-on-write
mechanism of the POSIX \code{fork}, the splitting process is very
fast, causing almost no delay in the execution.

We have evaluated our approach on eleven C programming projects with totally 337122 mutants and 20736 tests. 
The evaluation shows that, on top of the state-of-art approach for accelerating mutation analysis, our approach further accelerates the analysis, with a speedup of 2.56X on average. We have implemented these optimizations in our tool AccMut for C language.



\wdel{The rest of the paper is organized as follows. Section II introduces some related work of accelerating mutation analysis. Section III details approaches of AccMut. Next, in Section IV we describe the implementation of AccMut. Finally, we show the experiment results (Section V) and conclude (Section VI).}


\section{Related Work}

In general, the work for accelerating mutation analysis can
be divided into lossy approaches and lossless approaches~\cite{jia2011analysis}.

\subsection{Lossy Acceleration}
A typical lossy approach is weak mutation~\cite{howden1982weak}, where a mutation is assumed to be killed by a test if the mutated code changes the current system
state. In this way, we do not need to execute any mutant program but only
need to test at the mutated points. Recently, weak mutation has been
further accelerated by using split-stream execution, which forks new
threads on calling mutated method in Java
bytecode~\cite{durelli2012toward}. However, the results become
imprecise as a mutant changing the system state does not necessarily
violate the test. In other words, weak mutation only evaluates the
capability of the test suite to trigger the fault, but not whether the
triggered fault can be propagated to the output and the capability of
the test suite to capture the propagated faults. Other lossy 
approaches include randomly sampling the mutants~\cite{wong1995reducing}, clustering
the mutants and sampling a small number of mutants from each
cluster~\cite{ji2009novel}, mutant operator selection only adopts several sufficient operators~\cite{zhang2010operator}, select an efficient mutant subset~\cite{gopinath2016measuring,mirshokraie2015guided}, and select tests to execute~\cite{zhang2013faster}. Zhang et al.'s work utilizes machine learning to predict the mutant execution results without any execution~\cite{zhang2016predictive}. 
Different to lossy approaches, AccMut is a lossless approach, accelerating mutation testing without sacrificing precision.

\subsection{Lossless Acceleration}
\wdel{\subsubsection{Removing Redundant Computation}}
A main type of lossless approaches seek to reduce redundant computation in mutation analysis. 

Mutation schemata~\cite{untch1993mutation} can compile all mutants
into a single executable file at a time\wdel{, reducing the complexity of
total compiling time from $O(m)$ to $O(1)$}. \wdel{By the definition, m}Mutants are
slight syntactic transformations of the original program, so the most
parts of their code are duplicated. Mutation schemata can reduce redundance in compilation.
\wdel{Compiling these duplicated code
leads to redundant computation in compilation. Mutation schemata using
a global variable to control which mutant is activated and the
controller of test must initialize the global before running the test.
However, mutation schemata does not reduce the testing time.
Some widespread mutation analysis tools are based on
mutation schemata such as Javalanche~\cite{schuler2009javalanche}.}

Split-stream execution~\cite{tokumoto2016muvm,gopinath2016topsy,King:1991:FLS:116633.116640}, as mentioned in the
introduction, is the technique that reduces the redundant computation
before the first mutated statement. Split-stream execution is first
proposed by Offutt et al.~\cite{King:1991:FLS:116633.116640}, and then
explored by several researchers~\cite{gopinath2016topsy,tokumoto2016muvm}. The basic idea is to start all
mutants execution in one main process, and split the execution into
different processes when the first mutated statement is encountered.
As discussed, split-stream execution only reduces the redundant
computation before the first mutated statements while our approach can
reduce redundant computations after those.


Several approaches~\cite{offutt1994using,papadakis2015trivial} exist for
detecting the equivalence of mutants. Once an equivalent group is
detected, only one mutant in the group needs to be executed.
Just et al.~\cite{just2014efficient} take a step further to
detect equivalence of mutants with respect to one test. Two
inequivalent mutants may behave completely the same under one test
execution and thus we only need to execute one of them for the
test.
Compared to these approaches, our approach is more
fine-grained as we can reduce the redundancy in part of the test execution.
For example, suppose an expression $e$ in a loop is mutated.
Under test $t$, $e$ is evaluated 10 times, where the first 9 times
the mutant evaluates to the same value as the original expression,
but the 10th evaluation gives a different value. Using either
equivalent mutant detection or Just et al.'s approach, the mutant
has to be considered as different from the original program and will
be fully executed, while in our approach the execution before the
10th evaluation will be shared.
In other words, equivalent mutant detection considers
absolute equivalence between mutants, Just et al.'s approach considers the
equivalence between mutants modulo test input, while our approach considers the
equivalence between statements modulo the current state, which is more fine-grained.


\wdel{\subsubsection{Parallel Mutation Analysis}}
Some lossless approaches seek for parallel execution of mutation analysis. 
Approaches supporting different architectures have been explored, such as MIMD~\cite{offutt1992mutation} and SIMD~\cite{krauser1991high}.
If we view each test-mutant pair as an independent program, we can parallelize their execution on a MIMD machine. 
On the other hand, if we view each test as a different input data, we can parallelize the execution of different tests on one mutant on a SIMD machine.
\wdel{Gopinath et al. mentions parallelization of mutation analysis by forking mutants~\cite{gopinath2016topsy} while their work overlaps MuVM's first order mutation techniques.
Since AccMut forks executions, it naturally supports parallel execution. Because the input data and codes of mutation analysis are almost the same, AccMut can untiles the nature for further reduction.}

\wdel{\subsubsection{Removing Unnecessary Computation}}
\wdel{
Finally, recently Zhang et al.~\cite{zhang2013faster} propose to prioritize the tests for each mutation so that this mutation shall be killed quicker. 
In applications such as evaluating a test suite, when a test kills a mutant, we do not need to execute the rest of the tests for that mutant. 
As a result, the quicker the tests kill the mutant, the faster the mutation analysis. 
Nevertheless, in some applications of mutation analysis, all of the mutants need to be executed, such as fault location. Thus prioritizing of tests will not work in these applications. 
}


Finally, in the application of evaluating a test, several papers~\cite{zhang2012regression,zhang2013faster,just2012using} propose to prioritize the tests for each mutation so that this mutation shall be killed quicker. The works are orthogonal to ours and can be used together with AccMut.


\section{Basic Framework of AccMut}

\subsection{Overview}

\begin{figure*}[ht]
  \centering
  \includegraphics[width=0.9\textwidth]{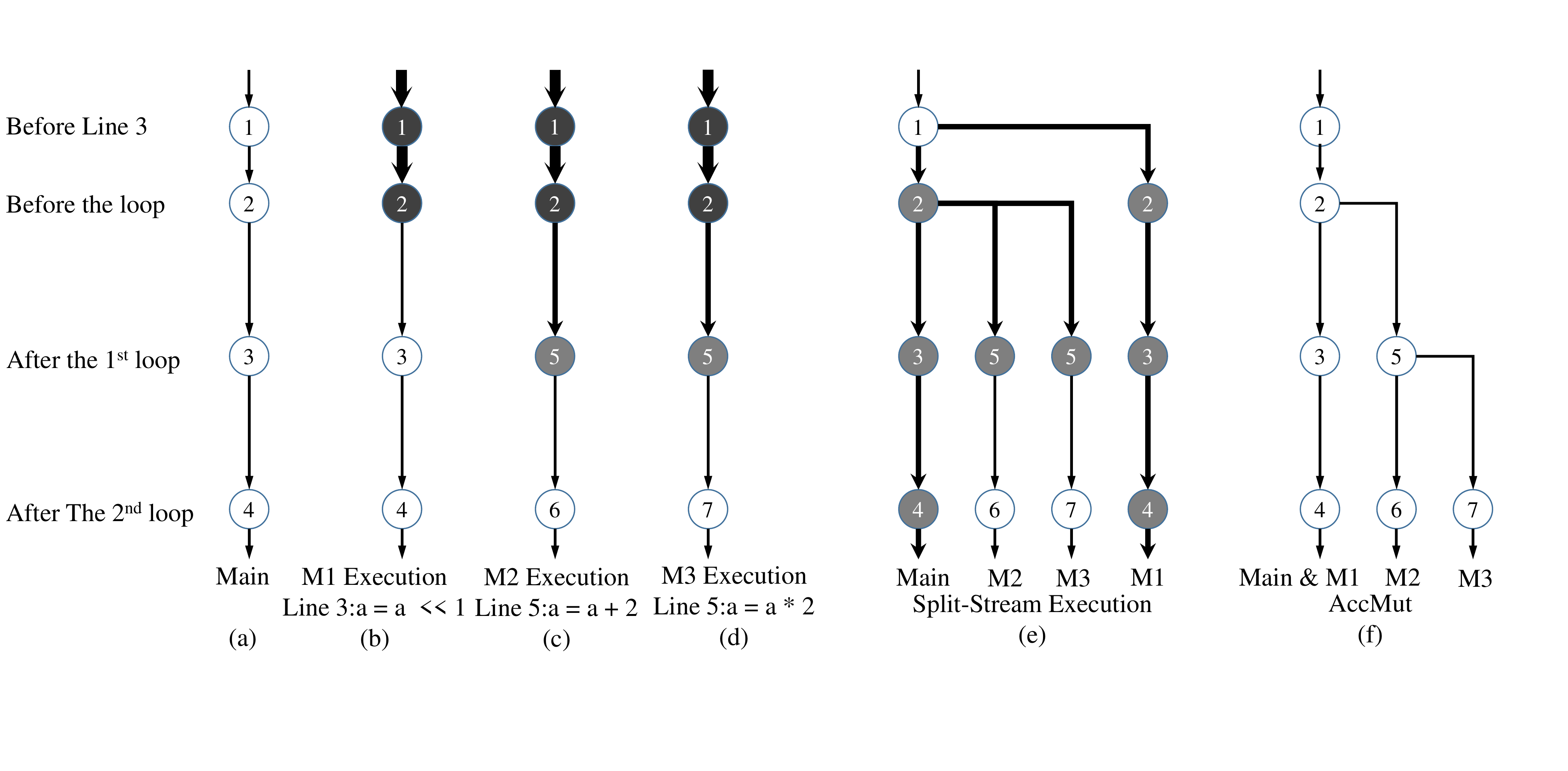}\vspace{-0.8cm}
  \caption{Dynamic Mutation Analysis with Equivalence
    Analysis\label{fig:redundancy2} }
\end{figure*}

We first describe the redundant execution that AccMut avoids with an
example in the following code snippet and in \figref{fig:redundancy2}. In  function \code{foo}, the line 6
is a computation intensive function without side-effects. The test
driver function is \code{test\_foo}, which sets the parameter \code{a} to \code{1}
and then judges the result. Let us assume that three mutants are generated in the function \code{foo}.
Mutant 1 (M1) changes \code{a = a + 1} at line 3 into \code{a = a << 1}.
Mutant 2 (M2) and Mutant 3 (M3) change \code{a = a / 2} at line 5 into \code{a = a + 2} and \code{a = a * 2}, respectively.

In standard mutation analysis, we execute all mutants for each test
and obtain their testing results. We show the execution of the three
mutants in \figref{fig:redundancy2}(b), (c) and (d), and the execution
of the original program in \figref{fig:redundancy2}(a) as a reference.

\lstset{numbers=left,numberstyle=\tiny,keywordstyle=\color{blue!70},commentstyle=\color{red!50!green!50!blue!50},frame=shadowbox, rulesepcolor=\color{red!20!green!20!blue!20},escapeinside=``,xleftmargin=2em,xrightmargin=2em, aboveskip=1em}

\begin{lstlisting}[language={[ANSI]C},numbers=left,basicstyle=\footnotesize, numberstyle=\tiny,keywordstyle=\color{blue!70},commentstyle=\color{red!50!green!50!blue!50},frame=none, rulesepcolor=\color{red!20!green!20!blue!20}]
int foo(int a){
  int i, res;
  a = a + 1; // M1:a << 1
  for(i = 0; i < 2; i++){
    a = a/2; // M2:a + 2, M3:a * 2
    res += time_consuming(a);
  }
  return res;
}
void test_foo(){
 assert(foo(1) == RESULT); 
}
\end{lstlisting}

The circles represent system states and the states with the same
number are the same. The states are captured at the program
  point listed on the left of \figref{fig:redundancy2}.
The arrows represent the transition of states after the executions of statements. 
And the length of the arrows means the execution effort during the transition.
If two transitions have the same starting state and the same ending
state, the two transitions are redundant. 
To show the redundancy between states and transitions, we use the brightness of the circles and the thickness of the arrows respectively.
The darker the circle is, or the thicker the arrow is, the more
redundant the state/transition is.
As we can see from \figref{fig:redundancy2}(b), (c) and (d), there are
several redundant transitions among the three mutant executions.
First, as the parameter $a$ of $foo$ is set to $1$ in this
test (state 1), the
results of $a = a + 1$ and $a = a << 1$ both equal to $2$
  (state 2). As a
result, the transitions before entering the loop, i.e., transitions to
state 1 and transitions between states 1 and 2, are
redundant among the three mutants. Second, during the first loop $a =
2$, the state of $a = a + 2$ and $a = a * 2$ are the same, and thus
the transition between states 2 and 5 in M2 and M3 is also redundant.
Note that this transition involves the call to a time-consuming
function, and thus the cost of the redundancy is high, so the length of the arrows during the loop is much longer. 


\figref{fig:redundancy2}(e) exhibits the execution of the mutants in
split-stream execution. In split-stream execution, all mutants start 
as one main process, and are later split from the main process at the first
mutated statements. The main process of our example is shown in the
first column in \figref{fig:redundancy2}(e). {M1 is split as a new process after state 1, and later M2
and M3 are split as new processes after state 2.}
{Note that we need to
keep the main process as more mutants may be split
from it.
As we can see, }Some of the redundancies are removed in
split-stream execution. For example, the transitions to state 1 is
shared among the three mutants as well as the main process.
However, two types of redundancies still exists. First, the transitions
between states 1 and 2 are the same between the main process and
M1, and thus it is not necessary to split M1 from the main
process. Second, although it is necessary to split M2 and M3 after
state 2, the transitions between state 2 and state 5, which involves
calling the time-consuming function, are still redundant among the two
split processes.


Our approach, AccMut, tries to further reduce the redundancies in
execution by exploiting the equivalence modulo the current state. An
execution of the three mutants in AccMut is shown in
\figref{fig:redundancy2}(f). Different from split-stream execution,
AccMut first classifies the next statements in different mutants into
equivalent classes modulo the current state, and uses one process to
represent each equivalent class. As a result, first, since the
mutated statement in M1 is equivalent to the original statement modulo
state 1, we would not split a new process for M1. Second, the two
mutated statements in M2 and M3 are equivalent modulo state 2, so we
split only one process for them. As a result, the redundant
transitions in \figref{fig:redundancy2}(e) are all removed in \figref{fig:redundancy2}(f).


More concretely, at each state of each process,
we conduct a trial execution of the statements and collect their changes to the system state. Then we cluster their changes to the system state into equivalence classes. 
If the number of equivalence classes is more than one, say $n$, we split $n-1$ new processes. 
Each forked process represents the mutants in one equivalence class, and we apply the change from the equivalence class to the state of the forked process.
Finally, the change from the remaining equivalent class is applied to the original process, and the original process now represents only the mutants in the remaining class.
This process continues for each process until all processes terminate.

However, in practice it may be expensive to store and cluster
the changes, specially when a statement makes a large set of changes.
For example, if a statement calls a procedure with a large side effects,
i.e., changing many memory locations, it may not be efficient to
record the changes of all memory locations and compare the changes
from different mutants. 

To solve this problem, in AccMut we record abstract changes rather
than concrete changes. An abstract change stores often much less
information than a concrete change, but nevertheless allows the
application of the change to the system state. For example, the change
of a procedure call can be represented abstractly by the address of
the procedure and all arguments passed to the procedure, rather than
all concrete changes produced by the procedure. When we need to apply
the change to system, we just actually invoke the method with the
arguments. In this way, we record only a small amount of information
allowing us to efficiently store and cluster the changes.

When two abstract changes are the same, applying them to the same
system state gives the same new state. However, the inverse is not
always true: when two abstract changes are different, they may not
always produce different states. For example, invoking a method with
different sets of arguments may not necessarily need to different
system state. In other words, the equivalence relation we computed is
conservative: when two statements are equivalent in the computed
equivalence relation, they are equivalent modulo the current state;
when two statements are inequivalent in the computed equivalence
relation, they may still be equivalent modulo the current state.

\wdel{
\revise{}{In addition, Just et al.'s approach~\cite{just2014efficient} needs a prepass to record all the related operands and a strong static analysis for loops and function invocations, and the cost of the prepass is not clear.}
}


\subsection{Definitions}

{In this sub section we define a set of concepts and operators that we
will use to describe our approach.}
{These definitions abstract away
concrete details in mutation operators such that our approach can be
generally applicable to different sets of mutation operators.}
Given a program, a mutation analysis first applies a set of mutation
operators to produce mutants from the program. Since the operations
can be applied in different granularities in different mutation
analyses, e.g., on instruction level, on expression level, or on statement level.
We use an abstract concept---location---to represent
the unit that a mutation operator applies. \wadd{Each mutant can be identified by a unique mutant ID.}
\wdel{Also we consider each
mutant can be identified by a unique mutant ID.}

\newcommand{\mids}{\ensuremath{I}\xspace}

{More concretely, a} program can be viewed as a set of locations. A
mutation procedure $p$ is a function mapping each location to a set of
variants. Each variant $v$ consists of a block of code (denoted as $v.code$) that can be
executed and a set of mutant IDs (denoted as $v.\mids$) that denote the mutants where this
variant of code block is enabled. The union of mutant IDs from all
variants at any two locations are the same, i.e., $\bigcup_{v \in p(l_1)} v.\mids=\bigcup_{v \in p(l_2)} v.\mids$ for
any mutation procedure $p$ and any two locations $l_1, l_2$, and the
union represents all mutant IDs in the system. Given any two variants from the same
location, their mutation IDs are disjoint, i.e., $v_1, v_2 \in p(l)
\Rightarrow v_1.\mids \cap v_2.\mids = \emptyset$ for any location $l$.
Given a mutation ID $i$, the code block $v.code$ at each
location $l$ where $i \in v.\mids \wedge v \in p(l)$ forms a new program, called a mutant. 

\wdel{
For example, let consider the following program with two lines of code.
\begin{verbatim}
1: a = a + 1;
2: b = b + 1;
\end{verbatim}
Suppose we have only one mutation operator: replace the basic
arithmetic operator (i.e., + - *) in a statement with another basic
arithmetic operator. Since this operator applies on statement level,
the above program has two locations: line~1 and line~2. At line~1,
this operator produces three variants: (a) \code{a = a + 1}, (b)
\code{a = a - 1}, and (c) \code{a = a * 1}, where (a) is the original
statement while (b) and (c) are mutated statements. Similarly, three
variants are produced at line~2: (d) \code{b = b + 1}, (e)
\code{b = b - 1}, and (f) \code{b = b * 1}. The four mutated
statements at lines~1 and 2 lead to four mutants, with ID 1-4. 
Then we have (a).\mids = \{3, 4\},
(b).\mids = \{1\}, (c).\mids = \{2\}, (d).\mids = \{1, 2\},
(e).\mids = \{3\}, (f).\mids = \{4\}.

This model also supports high-order mutants. If a high-order mutant is
created by mutating two statements in a program, the two mutants are
both enabled for the mutant, and at other locations the original
statements are enabled.
}

\newcommand{\execute}{\ensuremath{\mathtt{execute}}\xspace}
\newcommand{\try}{\ensuremath{\mathtt{try}}\xspace}
\newcommand{\apply}{\ensuremath{\mathtt{apply}}\xspace}

The execution of a mutant is represented by a sequence of system
states. A special function $\phi$ maps each system state to a
location, which indicates the next code block to execute. The
execution terminates at state $s$ when $\phi(s)=\bot$. Operation
\execute executes a code block on a system state. Given a variable
$s$ containing a system state and a code block $c$, $\mathit{\execute(s,c)}$
updates the system state in-place. Operation \execute can be
decomposed into two operations \try and \apply. Invocation
$\mathit{\try(s,c)}$ executes code block $c$ on system state $s$, and returns
a (potentially abstract) change $x$ describing the changes to the system state, without
actually changing $s$. Invocation $\mathit{\apply(x, s)}$ applies the change
$x$ in-place to a variable $s$ containing a system state. We require
that invoking $\apply(\try(s, c), s)$ is equivalent to invoking
$\execute(s, c)$. Please note that while $x=y \Rightarrow \apply(x,
s)=\apply(y, s)$ holds, $x\neq y \Rightarrow \apply(x,
s)\neq \apply(y, s)$ does not necessarily hold, allowing us to define
abstract changes.

\wdel{Note that
this execution model implies that, when there is a jump statement in any
mutant, it only jumps within its own location or to the start of
another location, but not the middle of another location. If such a
jump statement exists, we need to further divide this location into
two locations.}

\newcommand{\filterv}{\ensuremath{\mathtt{filter\_variants}}\xspace}
\newcommand{\filterm}{\ensuremath{\mathtt{filter\_mutants}}\xspace}
\newcommand{\filterom}{\ensuremath{\mathtt{filter\_out\_mutants}}\xspace}
\newcommand{\fork}{\ensuremath{\mathtt{fork}}\xspace}
\newcommand{\lchange}{\ensuremath{\mathtt{large\_change}}\xspace}

To implement AccMut efficiently, we also need
 three additional operations, as follows. The time complexity of the three
 operations should be as small as possible, preferably in constant time.
\begin{itemize}
\item \fork. Similar to the POSIX system call \code{fork()}, this
   operation splits a child process from the current process.
\item $\mathit{\filterv(V, I)}$. This operation filters a set of
variants in-place based on a set of mutant IDs, leaving only the variants
enabled for the mutants, i.e., $V$ is updated to $\left\{v\mid v\in V \wedge v.\mids
  \cap I \neq \emptyset \right\}$. The variants are assumed to be at the same location.
\item
  $\mathit{\filterm(I, V)}$.
  This operation filters a set of mutant IDs in-place based on a set of
  variants, leaving only the mutants
  containing one of the variant, i.e., $I$ is updated to $\left\{i\mid i\in I \wedge \exists v
    \in V. i \in v.\mids \right\}$. The variants are assumed to be at
  the same location.
\end{itemize}

\subsection{Standard Mutation Analysis}
Based on the definitions, we can build algorithms for mutation
analysis. We shall introduce AccMut step by step. We first start with
the standard mutation analysis without any acceleration (this
subsection), then we extend the algorithm into split-stream execution,
and finally we extend split-stream execution into AccMut.

The algorithm for standard
mutation analysis is as shown in \algref{alg:static}. 
Given all
mutant IDs in the system, the algorithm executes them one by one (line~1). The
execution of a mutant is a series of state transitions until there is
no code block to execute (line~3). At each
transition, the system first selects proper variant at the current
location (line~4), and then executes the variant (line~5).
Finally, necessary information about the execution result is recorded
by calling save($s,i$) (line~7).

Note that a direct implementation of the algorithm effectively
executes the program with an interpreter. Another way of
implementation is to apply $p$ beforehand, and instrument this
algorithm into the target program. This implementation is
equivalent to mutant schemata~\cite{untch1993mutation}, where all
mutants are generated into the program to save compilation cost.

\wdel{
\begin{algorithm}[t]
  \KwIn{$p$: a mutation procedure}
  \KwData{$s$: the current system state}
  \For{each mutant ID $i$ in all mutant IDs}{
    $s \leftarrow$ the initial system state\\
    \While{$\phi(s) \neq \bot$}{
      $\{v\} \leftarrow \filterv(p(\phi(s)), \{i\})$\\
      $\execute(v.code, s)$
    }
    {save}$(s, i)$
  }
\caption{Standard Mutation Analysis}
\label{alg:static}
\end{algorithm}
}

\newcommand{\filtervALG}{\ensuremath{\mathrm{filter\_variants}}\xspace}
\newcommand{\filtermALG}{\ensuremath{\mathrm{filter\_mutants}}\xspace}

\newcommand{\executeALG}{\ensuremath{\mathrm{execute}}\xspace}
\newcommand{\tryALG}{\ensuremath{\mathrm{try}}\xspace}
\newcommand{\applyALG}{\ensuremath{\mathrm{apply}}\xspace}

	\begin{algorithm}\small
		  \KwIn{$p$: a mutation procedure}
		  \KwData{$s$: the current system state}
		  \For{each mutant ID $i$ in all mutant IDs}{
		  	$s \leftarrow$ the initial system state\\
		  	\While{$\phi(s) \neq \bot$}{
		  		$\{v\} \leftarrow \filtervALG(p(\phi(s)), \{i\})$\\
		  		$\executeALG(v.code, s)$
		  	}
		  	{save}$(s, i)$
		  }
		  \caption{Standard Analysis}
		  \label{alg:static}
	\end{algorithm}

	\begin{algorithm}\small
		  \KwIn{$p$: a mutation procedure}
		  \KwData{$s$: the current system state}
		  \KwData{$I$: a set of mutant IDs represented by the current process}
		  $I \leftarrow$ all mutant IDs\\
		  $s \leftarrow$ the initial system state\\
		  \While{$\phi(s) \neq \bot$}{
		  	{proceed}$(p(\phi(s)))$
		  }
		  \For{$i \in I$}{
		  	{save}$(s, i)$
		  }  
		  \caption{Main Loop of Split-stream Execution and AccMut}
		  \label{alg:main}
	\end{algorithm}

\subsection{Mutation Analysis of Split-Stream Execution}
The main loop of split-stream execution is in \algref{alg:main}.
There are three differences from standard mutation analysis:
(1) for each process, there is a set $I$ indicating the mutant IDs
represented by the current process, which is initialized to all
mutant IDs; (2) at
the end of execution, save() is called for each mutation ID in $I$;
(3) a procedure proceed() is called for state transitions and
execution forking.

The procedure of proceed() is shown as \algref{alg:sseproceed}. If there is only one variant to execute,
we directly execute the variant and return (lines 2-6). 
If there are more than one variant, we first select one variant to be
represented by the current process (line 7), and then fork a new process for
each remaining variant (lines 8-15). When we fork a new process, the new process represents
the mutant IDs of the variants in the equivalence class (line 11), and
the corresponding variant is executed (line 12).
Finally, we update the mutant IDs of the original process and execute
the selected variant (line 16-17).

\begin{algorithm}[t]
	\KwIn{$V$: a set of variants at current location}
	\KwData{$s$: the current system state}
	\KwData{$I$: a set of mutation IDs represented by the current process}
	${\filtervALG}(V, I)$\\
	\If{$|V| = 1$}{
		$v \leftarrow$ the only variant in $V$\\
		\executeALG($v.code, s$)\\
		\Return
	}
	$v' \leftarrow$ a random variant in $V$\\
	\For{each $v$ in $V$ where $v \neq v'$}{
		$pid \leftarrow$ fork()\\
		\If{in child process}{
			$\filtermALG(I, \{v\})$\\
			\executeALG($v.code, s$)\\
			\Return{    //child process directly return}
		}
	}
	$\filtermALG(I, {v'})$\\
	\executeALG($v'.code, s$)\\
\caption{Algorithm of proceed($V$) in Split-stream Execution}
\label{alg:sseproceed}
\end{algorithm}

\subsection{Mutation Analysis in AccMut}
\wdel{\algref{alg:main} shows the main loop of mutation analysis in
  AccMut. There are three differences from standard mutation analysis:
  (1) for each process, there is a set $I$ indicating the mutant IDs
  represented by the current process, which is initialized to all
  mutant IDs; (2) at
  the end of execution, save() is called for each mutation ID in $I$;
  (3) a procedure proceed() is called for state transitions and
  execution forking. }

\wdel{
\begin{algorithm}[t]
  \KwIn{$p$: a mutation procedure}
  \KwData{$s$: the current system state}
  \KwData{$I$: a set of mutant IDs represented by the current process}
  $I \leftarrow$ all mutant IDs\\
  $s \leftarrow$ the initial system state\\
  \While{$\phi(s) \neq \bot$}{
    {proceed}$(p(\phi(s)))$
  }
  \For{$i \in I$}{
    {save}$(s, i)$
  }  
\caption{Main Loop of Mutation Analysis in AccMut}
\label{alg:main}
\end{algorithm}
}

The main loop of AccMut is the same as split-stream execution in \algref{alg:main}, however 
the algorithm of proceed() is different.
%
%
As shown
in \algref{alg:advanced}, 
first we will check the number of variants, which is the same as
split-stream execution (lines 2-6).
\revise{Otherwise, we
first try to execute the variants and cluster the changes to the
system state in equivalence classes (lines
7-11). Changes in each equivalence class will produce exactly the same
new system state when applied to the current system state. If there
are more than one equivalence classes, we fork a new process for each
extra equivalence class (lines 13-15). When we fork a new
process, the new process represents the mutant IDs of the variants in
the equivalence class (line~17)
, and we apply the change to the system state (lines 18-20). Finally, we
update the mutant IDs and system state for the current process (lines 25-26).
}{
  The main difference starts from lines 7-10, where we first collect
  the changes produced by the mutants into set $X$. Then we cluster the
  changes into equivalence classes (line 11). The rest of the
  algorithm has a similar structure to split-stream execution, where
  each equivalent class corresponds to a variant in split-stream
  execution. We first select a class that the current process
  represents (line 12), and then fork a new process for each other
  cluster (lines 13-22), and
  finally update the mutant IDs and the state of the current process
  (lines 23-26). 
}

\wdel{
\revise{}{Please note that in general statements have side effect and
  a code block may perform a large amount of change. In such a case,
  it may not be efficient to store and group the changes. In our
  implementation, we detect such cases statically by the type of a mutant and directly fork a
  new process for each variant without \code{try} and grouping, as in
  split-stream execution.}
}

\begin{algorithm}[t]
  \KwIn{$V$: a set of variants at current location}
  \KwData{$s$: the current system state}
  \KwData{$I$: a set of mutation IDs represented by the current process}
  ${\filtervALG}(V, I)$\\
  \If{$|V| = 1$}{
    $v \leftarrow$ the only variant in $V$\\
    \executeALG($v.code, s$)\\
    \Return
  }
  $X = \emptyset$\\
  \For{each $v$ in $V$}{
          $X \leftarrow X \cup \{\tryALG(v.code, s)\}$
  }
  $\mathbb{X} \leftarrow$ group changes in $X$ into equivalent classes\\
  $X_{cur} \leftarrow$ any one of the equivalent classes in $\mathbb{X}$\\
  \For{each equivalence class $X$ in $\mathbb{X} - \{X_{cur}\}$}{
    $V \leftarrow$ the variants corresponding to changes in $X$\\
    $pid \leftarrow$ fork()\\
    \If{in child process}{
      $\filtermALG(I, V)$\\
      $x \leftarrow$ a random change in the equivalence class $X$\\
      $\applyALG(x, s)$\\
      \Return{}
    } 
  }
  $V \leftarrow$ the variants corresponding to changes in $X_{cur}$\\
  $\filtermALG(I, V)$\\
  $x \leftarrow$ a random change in $X_{cur}$\\
  $\applyALG(x, s)$
\caption{Algorithm of proceed($V$) in AccMut}
\label{alg:advanced}
\end{algorithm}



{
\subsection{Correctness of AccMut}

\begin{theorem}
  The algorithms of standard mutation analysis, split-stream
  execution, and AccMut produce exactly the same sequence of invocations
  to $save(s, i)$ with the same arguments. 
\end{theorem}
    \trivlist
    \item[%
         \hskip 10pt
         \hskip \labelsep
        {\sc Proof Sketch.}%
    ]
    \ignorespaces
  This can be proved by an induction over the number of state
  transitions to show that the same sequence of
  state transitions occurs for each mutant in all three algorithms. When
  the length is zero, this property trivially holds. When length is
  $k$, we can see that both algorithm selects the same variant to
  execute for each mutant, and thus the property holds for length
  $k+1$.
    \endtrivlist

}

\revise{
\section{Realizing AccMut}
In this section we discuss how to realize AccMut for specific
mutation analyses. To realize AccMut, we need to provide efficient
implementation for the three operations: $\mathit{\fork}$, $\mathit{\filterv(V, I)}$, and
$\mathit{\filterm(I, V)}$. We shall discuss how to implement fork() efficiently
on POSIX systems, and how to implement $\mathit{\filterv(V, I)}$, and
$\mathit{\filterm(I, V)}$ efficiently for first-order mutation analysis. 
}{\section{Implementing AccMut for First-Order Mutation Analysis on
  LLVM IR}
  Previous section gives the basic framework of our algorithm. In this
  section we demonstrate how to implement this framework for
  first-order mutations on LLVM IR. LLVM~\cite{lattner2004llvm} is a widely used compiler
  framework where different front-ends exist to translate different
  high-level languages such as C, C++, Java, into its intermediate
  representation (IR).
  To implement the framework, we need to design the
  mutation operators, and implement operations including \try,
  \apply, \fork, \lchange, \filterv, and
  \filterm.}

  Our implementation is available online\footnote{
  \url{https://www.dropbox.com/s/9zc00frihqe5elb/accmut.zip?dl=0}
  (This is a temporary address for double-blinded review).}.

\wdel{
\subsection{Overall structure of AccMut}
	AccMut contains three main components: Mutation Generator, Mutation Instrumenter and Runtime Library. AccMut processes mutation analysis by the following steps.
	\begin{enumerate}[1)]
		\item  Compile the source code of the program into LLVM IR. Mutation Generator scans the IR code and generates the description file of mutants on the LLVM IR code.
		\item  Mutation Instrumenter instrument mutants into the IR code of the program using mutation schemata. Then compile the IR code into object files.
		\item  LLVM links the object files with Runtime Library and obtains an executable file of the program.
		\item  The executable file executes each test on all mutants.
	\end{enumerate}
}

\subsection{Mutation Operators}
  As we mutate on the LLVM IR level, each IR instruction corresponds
  to a location. 
  \tblref{tab:operators} describe the
  mutation operators considered in our implementation. These operators are
  designed by mimicking operators on existing IR-based mutation tools.
  Since we do not find any mutation tool working on LLVM IR code, we
  mimic the operators from two mutation tools that work on Java byte
  code, which takes a similar form to LLVM IR. The two tools we chose
  are Major~\cite{just2011major,Just:2014:MMF:2610384.2628053} and
  Javalanche~\cite{schuler2009javalanche}, both of which are widely
  used mutation tools for Java. We converted all operators from Major
  and Javalanche except for two operators from Javalanche manipulating
  Java concurrency classes such as \code{Monitor}. These
  operators cannot be converted because there is no corresponding
  instruction or functions in LLVM IR level.
  
\begin{table}[t]
	\centering
	\caption{Mutation Operators}
	\label{tab:operators}
	\begin{tabular}{|c|c|c|}
		\hline
		Name & Description & Example \\
		\hline
		AOR & Replace arithmetic operator & $a~+~b$ $\rightarrow$ $a~-~b$\\
		LOR & Replace logic operator & $a~\&~b$ $\rightarrow$ $a~|~b$\\
		ROR & Replace relational operator & $a==b$ $\rightarrow$ $a>=b$\\ 
		LVR & Replace literal value & $T$ $\rightarrow$ $T+1$\\
		COR & Replace bit operator & $a~\&\&~b$ $\rightarrow$ $a~||~b$\\
		SOR & Replace shift operator & $a>>b$ $\rightarrow$ $a<<b$\\
		STDC & Delete a call & $foo()$ $\rightarrow$ $nop$\\
		STDS & Delete a store & $a=5$ $\rightarrow$ $nop$\\
		UOI & Insert a unary operation & $b=a$ $\rightarrow$ $a++;~b=a$\\
		ROV & Replace the operation value & $foo(a,b)$ $\rightarrow$ $foo(b,a)$\\
		ABV & Take absolute value& $foo(a,b)$ $\rightarrow$ $foo(abs(a),b)$\\
		\hline
	\end{tabular}
\end{table}

\subsection{Implementing \fork}
We implement operation \code{fork} using the POSIX system call
\code{fork}. The system call uses a copy-on-write mechanism~\cite{architecturebook},
where the newly created process shares the same memory space as the
original process, and only when a process performed a writing operation,
the corresponding memory page is copied and stops to be shared between
processes. In this way, the \code{fork}
operation can be completed in constant time.

However, the system call only enables the copy-on-write mechanism for virtual memory
access and does not support IO operations. In our implementation we
implemented a similar copy-on-write mechanism for file access by
replacing the original IO functions in the C library. 
Basically, files are divided into pages and a page table is maintained
for the positions of pages, and all file accesses are redirected to the
query of the page table.
When a forked process wrote to a shared fragment, the fragment is
copied to a new file and the page table in the process is updated. 

\subsection{Implementing \try and \apply}
The key point of implementing \try and \apply is to define the format
of the abstract changes. 
Since LLVM IR is in the three-address form, most instructions modify
only one memory location with primitive value. For those instructions, the
changes are easy to define: we just need to record the memory location
the instruction modifies and the new value the instruction stores at
the location.

The only exception is procedure call, as the callee procedure may
modify many different locations. As mentioned in the overview section,
the abstract change by a procedure call is defined as a tuple
including the address of the procedure and the arguments passed to the
procedure. Please note that in LLVM IR all passed arguments are
well typed. For example, an invocation of the function
$\mathit{void~foo}(int, int)$ has two mutants, an ROV
($\mathit{foo}(a,b)$$\rightarrow$$\mathit{foo}(b,a)$) and an STDC (remove the call). The
three variants give three changes, $\mathit{foo}(a, b)$, $\mathit{foo}(b, a)$, and an
empty change. The first two changes are equivalent only when $a=b$,
and the third change is always different from the first two.

\subsection{Implementing filter\_variants and filter\_mutants}

Since \filterv will be performed at every
location, and \filterm will be performed every time we fork a
process, it is better to keep the time complexity of the two
operations small, preferable $O(1)$. In this sub section we discuss how to implement the two
	operations for first-order mutation.

The challenges of implementing the two operations is that both operations
require the computation of set intersection, and a standard set
implementation has a time complexity of $O(n\log{n})$, where $n$ is
the set size. Since the set may contain all mutant IDs, this method
is too costly.
To get an efficient implementation, we utilize the fact that the
number of variants at each location has a small upper bound $u$. If
the complexities of the operations only depend on $u$ but not the
total number of mutants, the complexity is $O(1)$. In other words, we
assume that $O(u)=O(u^2)=O(\log{u})=\ldots=O(1)$.

More concretely, at each location, there are two types of variants. First,
variants generated by applying a mutation operator on the current
the location, called \emph{mutant variants}. Such a variant is always enabled for only one mutant ID.
Second, the variant of the original instruction, called \emph{original
variant}. The variant is
enabled for all remaining mutant IDs.

Utilizing this fact, we can design different data structures for
different sets. First, for each process, there is a set of mutant IDs
that the current process represents. Initially the set contains all
mutant IDs, but each time a process is forked into a set of processes,
there is at most one process executing the original variant and the
sizes of mutation IDs in all other processes are smaller than $u$.
Therefore, we use two different data structures to represent sets of
mutation IDs. The mutation IDs of the initial process is represented
by a bit vector. Each bit corresponds to a mutant ID, where one
indicates this mutant ID is in the set, and zero indicates this mutant
ID is not in the set. With a bit vector, operations such as adding,
removing, membership query can be finished in $O(1)$ time. Whenever a
process is forked, the process executing the original variant
inherits the bit vector from its parent, and for all other processes,
we create a new linked list for storing the mutant IDs. Since the size
of the list is bounded by $u$, the operations on the list, such as
membership query, is effectively $O(1)$.


\newcommand{\ori}{{\tt ori\_variant}\xspace}
\newcommand{\mut}{{\tt mut\_variants}\xspace}
\newcommand{\inc}{{\tt ori\_included}\xspace}
Second, there is a variable $V$ in
Algorithm~\ref{alg:advanced} storing a set of variants. Also for
each variant, there is a set of mutant IDs that the variant
represents. We represents these sets by treating the two types of
variants different. We use a data structure {\tt VariantSet} to store
a set of variants. More concretely, a {\tt VariantSet} is a tuple,
$(\ori, \inc, \mut)$, where \ori is the code of the original
variant, \inc is a Boolean variable indicating whether the original
variant is included in the set or not, and \mut is a linked list of mutant
variants. Each mutant variant $v$ consists of a block of code and one
mutant ID. To avoid confusion, we shall use $v.i$ to
represent the only mutant ID. In this way, we can quickly test whether
the original variant is included in the set or not. Also the size of
\mut is bound by $u$, so the operations on the set is effectively
constant time.

The algorithm for implement \filterv is shown in \algref{alg:filterv}.
We first filter the mutant variants (lines 3-6).
Since all operations have the complexity of $O(1)$ (note that $O(u)=O(1)$) and the loop is
bounded by $u$, this part has the complexity of $O(1)$. 
Next, we
consider whether the original variant should be filtered out
or be kept (lines 8-12). Since each mutant variant is enabled
for one mutant, if the currently selected variants are fewer than the
mutants represented by the current process, there must be remaining mutants
and the original variant should
also be selected. It is easy to see this part also takes $O(1)$. As a
result, the complexity of \filterv is $O(1)$.

The algorithm of \filterm is shown in \algref{alg:filterm}. 
If the original mutant is included in $V$, we build
the result negatively by removing mutant IDs (lines 1-4), otherwise we
build the result positively by adding mutant IDs (lines 6-9). Since
all operations and all loops are bound by $u$, the whole algorithm takes $O(1)$.

\begin{algorithm}
	\KwIn{$V$: a set of variants to be filtered}
	\KwIn{$I$: a set of mutation IDs used to filter $V$}
	$V' \leftarrow $ a new {\tt VariantSet}\\
        $V'.\ori \leftarrow V.\ori$\\
	\For{each $v \in V.\mut$}{
		\If{$I.contains(v.i)$}{
			$V'.\mut.add(v)$
		}
	}
	\eIf{$V'.\mut.size < I.size$}{
          $V'.\inc=true$
        }{
          $V'.\inc=false$
        }
	$V \leftarrow V'$  
	\caption{\filterv}
	\label{alg:filterv}
\end{algorithm}

\begin{algorithm}
  \KwIn{$I$: a set of mutation IDs to be filtered}
  \KwIn{$V$: a set of variants used to filter $I$}
  \KwData{$MV$: all mutant variants at the current location}
  \eIf{$V.\inc$}{
    \For{each $v \in MV - V.\mut$}{$I.remove(v.i)$}
  }
  {
    $I \leftarrow$ a new empty linked list\\
    \For{each $v \in V.\mut$}{
      $I.add(v.i)$
    }
  }
  \caption{\filterm}
  \label{alg:filterm}
\end{algorithm}

\subsection{Parallelism Control} A delicate point in our implementation is parallelism control. If a
large number of mutants can be generated from a program, we may fork a
large number of processes. Too many processes may lead to a large
overhead in scheduling these processes\wdel{, regrading the performance of
whole analysis}. As a result, we limit the number of parallel processes
in the implementation. In our current implementation we limit the
number of parallel processes to one. There are two reasons for this
limit. (1) In the evaluation shown later, this design controls the
uncertainty brought by parallel execution, giving a more stable
execution time. (2) The parallelism management is simpler: each time
we fork a child process, we suspend the parent until the child process
exits. Furthermore, though we only allow one parallel process for a
test execution, it is still possible to parallel mutation analysis
using our tool: we can parallelize the executions of different tests.


\revise{\section{Tool Implementation}
We have implemented our approach as a tool for first-order C mutation analysis.
Our tool is built on top of LLVM, a widely-used compiler framework.
The front-end of LLVM, such as Clang, first translates the high-level programming language code
into intermediate representation (IR). Then the back-end compiles the IR code into executables.
To make our implementation extensible, our tool performs mutations at the IR
level, such that our tool can be easily exported to other languages.}{}


\revise{As we mutate on the LLVM IR level, each IR instruction corresponds to
a location. We apply a set of mutation operators on IR
instructions to produce mutants.
\tblref{tab:operators} describe the
mutation operators used in our implementation. These operators are
designed by mimicking operators on existing IR-based mutation tools.
Since we do not find any mutation tool working on LLVM IR code, we
mimic the operators from two mutation tools that work on Java byte
code, which takes a similar form to LLVM IR. The two tools we chose
are Major~\cite{just2011major,Just:2014:MMF:2610384.2628053} and
Javalanche~\cite{schuler2009javalanche}, both of which are widely used
mutation tools for Java. We converted all operators from Major and
Javalanche except for two operators from Javalanche manipulating Java
concurrency classes such as \code{Monitor}.{ These operators cannot be
converted because there is no corresponding instruction or functions
in LLVM IR or in the C language.}}{}

\revise{A delicate point in our implementation is parallelism control. If a
large number of mutants can be generated from a program, we may fork a
large number of processes. Too many processes may lead to a large
overhead in scheduling these processes\wdel{, regrading the performance of
whole analysis}. As a result, we limit the number of parallel processes
in the implementation. In our current implementation we limit the
number of parallel processes to one. There are two reasons for this
limit. (1) In the evaluation shown later, this design controls the
uncertainty brought by parallel execution, giving a more stable
execution time. (2) The parallelism management is simpler: each time
we fork a child process, we suspend the parent until the child process
exits. Furthermore, though we only allow one parallel process for a
test execution, it is still possible to parallel mutation analysis
using our tool: we can parallelize the executions of different tests.}{}



\revise{Our implementations and all experimental data are open\footnote{	{Hide the homepage for the double-blind review}
  }.}{}


\section{Evaluation}
{Our evaluation aims to answer the research question: How does AccMut perform compared to existing approaches?}





\begin{table*}[t]
	\centering
	\caption{Summary of the Recent Work on Mutation Analysis}
	\label{tab:summary}
	\begin{tabular}{|c|c|c|c|c|c|c|c|}
		\hline
		Author(s) [Reference]  & Conference & Language & Largest Subject & Max Mutant Number & Sampled Mutants\\
		\hline
		Just et al.~\cite{just2012using} & ISSRE'12 & Java & 116,750 & 160,891 & No \\
		Gligoric et al.~\cite{gligoric2013selective} & ISSTA'13 & Java & 68,218 & 738 & No \\
		Zhang et al.~\cite{zhang2013faster} & ISSTA'13 & Java & 36,910 & 36,418 & Yes \\
		Harman et al.~\cite{harman2014angels} & ASE'14 & Java & 6,359 & 7,204 & Yes \\
		Just et al.~\cite{just2014efficient} & ISSTA'14 & Java & 116,750 & 108,174 & No \\
		Yao et al.~\cite{yao2014study} & ICSE'14 & C & 35,545  & 3,838  & Yes\\
		Papadakis et al.~\cite{papadakis2015trivial} & ICSE'15 & C & 362,769 (39,008) & 72,432  & Yes\\
		Kurtz et al.~\cite{kurtz2016analyzing} & FSE'16 & C & 512 & 11,080  & Yes\\
		Gopinath et al.~\cite{gopinath2016limits} & ICSE'16 & Java & Not given & 122,484 & Yes \\
		Zhang et al.~\cite{zhang2016predictive} & ISSTA'16 & Java & 29,702 & 28,133 & Yes \\
		Papadakis et al.~\cite{papadakis2016threats} & ISSTA'16 & C & 20,461 & 22,023 & Yes \\		
		\hline
		This paper & --- & C & 477,257 (42,073) & 173,683 & No \\
		\hline
	\end{tabular}
	{\small
		
	}
\end{table*}

\begin{table*}[t]
	\centering
	\caption{Subject Programs}
	\label{tab:subjects}
	\begin{tabular}{|c|c|c|c|c|c|c|c|}
		\hline
		Name  & LOC & Tests & Mutants & Locations & Avg $u$  & Max $u$ & Description\\
		\hline
		flex & 10334 & 42 & 56916 & 5119 & 11.1  & 32 & a lexical analyzer generator\\
		gzip & 4331 & 214 & 37326 & 3058 & 12.2  & 22 & a tool for file compression\\
		grep & 10102 & 75 & 58571 & 4373 & 13.4 & 34 & a tool for searching plain-text files\\
		printtokens & 475 & 4130 & 1862 & 199 & 9.4 & 22 & a lexical analyzer\\
		printtokens2 & 401 & 4115 & 2501 & 207 & 12.1 & 22 & an other lexical analyzer\\
		replace & 512 & 5542 & 3000 & 220 & 13.6 & 22 & a tool for pattern matching\\
		schedule  & 292 & 2650 & 493 & 55 & 9.0 & 22 & a priority scheduler\\
		schedule2 & 297 & 2710 & 1077 & 121 & 8.9 & 22 & another priority scheduler\\
		tcas & 135 & 1608 & 937 & 73 & 12.8 & 35 & an aircraft collision avoidance tool\\
		totinfo & 346 & 1052 & 756 & 63 & 12.0 & 22 & a statistics tool\\
		vim 7.4 & 477257 (42073) & 98 & 173683 & 14124 & 12.3 & 43 & a text editor\\		
		\hline
		Total &  504482 & 20736 & 337122 & 27612 & 12.2 & --- & ---\\
		\hline
	\end{tabular}
	{\small
	}
\end{table*}

\subsection{Subjects}
To answer the research questions,  we totally collected eleven subjects.
The statistics of the selected programs are shown in
\tblref{tab:subjects}. Ten of the subjects are from the SIR
repository~\cite{sir} and one (\emph{vim 7.4}) is from an open source
project. We select these subjects because they cover a wide range of
application domains from GUI editor to command-line compression, cover both
real-world applications and widely-used benchmarks, and cover
applications of different sizes from hundreds of lines to hundreds of
thousands of lines. Also, many of the subjects have been used in
existing studies. For example, \emph{vim 7.4} is the new version of the
largest subject (\emph{vim 7.2}) used in Papadakis et al.'s empirical study~\cite{papadakis2015trivial}, which is
one of the largest empirical study on mutation analysis in recent
years. Please note that \emph{vim 7.4} has about 11k more lines of code
than \emph{vim 7.2}, making our subject even larger. Since the whole
\emph{vim} is too large for evaluation, following Papadakis et
al.~\cite{papadakis2015trivial}, we selected the largest two
components, \emph{eval} and \emph{spell}, as targets of the mutation operators. In \tblref{tab:subjects}, we list the LOC of the two
components in parentheses.

The subjects have total 504482 lines of code, 20736 tests and 337122
mutants upon 27612 IR-level locations.\footnote{LOC is collected by
  the tool \code{cloc} (\url{http://cloc.sourceforge.net}).} The
average of $u$, namely the average number of mutants per mutated location, is
12.2. The max $u$ of a subject is in the range of 22 to 43. Thus,
\algref{alg:filterv} and \algref{alg:filterm} can be bounded in
constant complexity generally.

Please note that the subjects we use are among the largest in studies
of mutation analysis. \tblref{tab:summary} shows the studies of mutation
analysis in recent years. As we can see from the table, our largest subject
is among the largest in terms of both the lines of code and the
number of mutants. Especially, the number of mutants is the highest
among all subjects. This is because other studies either use fewer
mutation operators, use more coarse-grained operators, or perform
sampling on the mutants, which reduce the number of mutants.

\wdel{
The subjects include large projects as well as small programs, and their lines of code amount to 21558, as shown in the second column of \tblref{tab:subjects}\wdel{\footnote{LOC is collected by \emph{cloc} only considering C source files and C header files.}}. 
}

\wdel{Also, the programs are designed for various areas: 
tcas is an aircraft collision avoidance system; 
printtoken is a lexical analyzers; 
flex is a fast lexical analyser generator;
grep is a command-line utility for searching plain-text data sets for lines matching a regular expression; 
replace performs pattern matching and substitution.

These subjects come with in total 11397 tests, as shown in the third column of \tblref{tab:subjects}. 
We employed all of the 11 mutation operators in \tblref{tab:operators} and generated in total 119176 mutants, as shown in the last column of \tblref{tab:subjects}.
}

\subsection{Procedures}
In the experiments we compared AccMut with two controlling techniques:
mutant schemata~\cite{untch1993mutation} and split-stream
execution~\cite{King:1991:FLS:116633.116640,gopinath2016topsy,tokumoto2016muvm}.
\wdel{As introduced before, mutant schemata reduces the redundancies in
compilation; split-stream execution is based on mutant schemata,
reducing the redundancies in execution before the first mutated
statement; our approach is further built upon split-stream execution
and reduces more redundancies in execution.}
However, we cannot find a publicly-available tool that implements both
mutant schemata and split-stream execution for C programs. The only
state-of-the-art tool that implements split-stream execution on C
within our knowledge is MuVM~\cite{tokumoto2016muvm}. However, this
tool\wdel{ is an internal tool for a company and } is not publicly available.
As a result, we implemented the two techniques by modifying the
implementation of our tool. Mutant schemata is implemented as
Algorithm~\ref{alg:static} and split-stream execution is implemented
as Algorithm~\ref{alg:main} and Algorithm~\ref{alg:sseproceed}.




\revise{We then executed the three techniques on our subjects and recorded the analysis time per each subject and each technique.}{}
In our experiment, we sequentially executed the tests but not parallelized them, in order to obtain more stable results.
Since there is at most one forked process running at a time, the
execution of mutants was also sequential.
Furthermore, we executed each subject using each technique three times and recorded the median execution time.

The experiments are performed on a laptop with 2.50 GHz Intel i7-4710MQ CPU and
12GB memory. The operating system is Ubuntu 16.04 LTS.

\subsection{Results}
The execution time of the three techniques is shown in
\tblref{tab:results}. From the table, we can make the following
observations.

\wdel{	
First, the testing results of the three techniques are identical,
which serves as an indication that the three techniques are likely to
be implemented correctly.}

\begin{table*}[t]
	\centering
	\caption{Experimental Results}
	\label{tab:results}
	\begin{tabular}{|c|c|c|c|c|c|c|}
		\hline
		Subjects & AccMut & SSE & MS & SSE/AccMut & MS/AccMut & MS/SSE\\
		\hline
		flex & 12m58s & 40m22s & 1h26m17s & 3.11x & 6.65x & 2.13x\\
		gzip & 50.4s & 2m32s & 55m19s & 3.02x & 65.85x & 21.84x\\
		grep & 2m19s & 7m16s & 58m56s & 3.13x & 25.36x & 8.10x\\
		printtokens & 11m55s & 23m36s & 2h10m54s & 1.94x & 10.98x & 5.67x\\
		printtokens2 & 11m35s & 38m7s & 57m24s & 3.30x & 4.97x & 1.51x\\
		replace & 17m27s & 41m22s & 44m56s & 2.37x & 2.57x & 1.09x\\
		schedule & 3m14s & 6m20s & 8m14s & 1.96x & 2.54x & 1.30x\\
		schedule2 & 6m40s & 13m18s & 17m05s & 2.00x & 2.56x & 1.28x\\
		tcas & 7.7s & 21.2s & 16m31s & 2.6x & 128.7x & 46.7x\\
		totinfo & 1m55s & 4m28s & 6m19s & 2.33x & 3.30x & 1.41x\\
		vim 7.4 & 1m9s & 2m10s & 3h26m6s & 1.88x & 179.2x & 95.1x\\
		\hline
		Total & 1h10m10s & 2h59m52s & 10h28m1s & 2.56x & 8.95x & 3.49x\\				
		\hline
	\end{tabular}\\
        {\small
		AccMut = Our approach, SSE = Split-Stream Execution~\cite{tokumoto2016muvm},
		MS = Mutant Schemata~\cite{untch1993mutation}, XXX/YYY = Speed up of YYY
		over XXX }
\end{table*}

\begin{table*}[t]
	\centering
	\caption{The Average Number of Executed Mutants For Each Test}
	\label{tab:executedPros}
	\begin{tabular}{|c|c|c|c|c|c|c|}
		\hline
		Subjects & AccMut & SSE & MS & AccMut/SSE & AccMut/MS & SSE/MS \\
		\hline
		flex & 5881.1 & 16610.7 & 56916 & 35.4\% & 10.3\% & 29.1\% \\
		gzip & 434.5 & 1638.1 & 37326 & 26.5\% & 1.2\% & 4.4\% \\
		grep & 1303.2 & 4014.4 & 58571 & 32.5\% &  2.2 \% & 6.9\% \\
		printtokens & 413.5 & 1019.3 & 1862 & 40.6\% & 22.2\% & 54.7\% \\
		printtokens2 & 750.4 & 1724.4 & 2501 & 43.5\% & 30.0\% & 68.9\% \\
		replace & 483.7 & 1484.1 & 3000 & 32.6\% & 16.1\% & 49.5\%\\
		schedule & 180.0 &  405.9 & 493 & 44.3\% & 36.5\% & 82.3\%\\
		schedule2 & 384.9 & 844.2 & 1077 & 45.6\% & 35.7\% & 78.4\%\\		
		tcas & 99.2 & 434.1 &  937 & 22.9\% & 10.6\%  &  46.3\% \\
		totinfo & 220.5 & 566.0 & 756 & 39.0\% & 29.2\% & 74.9\%\\
		vim 7.4 & 601.0 & 1472.7 & 173683 & 40.8\% & 0.3\% & 0.8\%\\
		\hline
		Average & 977.5 &  2746.7 & 30647.4 & 35.6\% & 3.2\% & 9.0\% \\
		\hline
	\end{tabular}\\
\end{table*}


\begin{itemize}
\wdel{
\item All the three techniques can finish mutation analysis on all
  the subjects within a certain amount of time. Among them, the
  analysis time of any subject in AccMut is smaller than 20 minutes,
  while mutant schemata can takes up to more than 3 hours.
}

\item On all subjects AccMut constantly outperforms the other
  two techniques, suggesting that exploiting the equivalence modulo a
  state can reduce the more redundant computation over the
  state-of-the-art techniques, and the benefit outperforms the extra overhead.
\item \wdel{On some subjects such as grep, the speedup can be up to 3.13x
  over split-stream execution, and be up to 25.36x over mutant
  schemata. On average, AccMut is 2.52x faster than split-stream
  execution. In other words, our approach can significantly boost the
  performance of mutation analysis over existing approaches.}
	\wadd{
		On average, AccMut is 2.56x faster than split-stream
		execution and 8.95x than mutation schemata. Our approach can significantly boost the
		performance of mutation analysis over existing approaches.
	}
\item Split-stream execution also significantly outperforms mutant
schemata, with an average speedup of 3.49x. This result is
consistent with an existing study~\cite{tokumoto2016muvm}.	
\end{itemize}

\revise{}{
\subsection{Detailed Analysis}
To further understand how AccMut
achieved the speedups, we performed a detailed analysis of the
execution process. First, the main reason our approach outperforms previous
approaches is that fewer mutants are executed, and we analyzed how
significant the reduction is. \tblref{tab:executedPros} shows the
average number of executed mutants for each approach. Mutation
schemata executes all mutants. Split-stream execution executed only
mutants covered by a test. AccMut executes much fewer mutants than
both approaches, as many covered mutants are still equivalent modulo
state.
}

\begin{table*}[t]
	\centering
	\caption{The Number of Executed Instructions}
	\label{tab:profile}
	\begin{tabular}{|c|c|c|c|c|c|c|}
		\hline
		Instruction Type & AccMut & SSE & MS & AccMut/SSE & AccMut/MS & SSE/MS \\
		\hline
		Original Program & 1,054,174 & 2,404,487 &   37,617,433 & 43.8\% & 2.9\%  &  6.4\% \\
		Extra Cost & 83,148,833 & 96,490,502 &  177,988,701 & 86.2\% & 46.7\%  &  54.2\% \\
		\hline
		Total Executed & 84,203,007 & 98,894,989 & 215,606,134 & 85.1\% & 39.1\% & 45.9\% \\
		\hline
	\end{tabular}\\
\end{table*}

\revise{}{
Besides reducing the mutants executed, AccMut may also introduce extra
overheads in trial execution of the instructions, clustering the
changes, etc. To further understand how the
introduced overheads compared to the saved executions, we measured the
number of the original instructions executed and the number of the
extra instructions executed in the three approaches. Note that a
better measurement here is the time used to execute the instructions, but the
LLVM profiler does not work on our implementation because of a signal
conflict, so we use the number of instructions to approximate the
time. Because of the high cost of tracing the instructions, we only
measured the first 100 tests of \emph{tcas} and \emph{printtokens}.

The result is shown in \tblref{tab:profile}. 
As we need to select a variant at each location, the extra
instructions executed is much more than the original instructions in
all three approaches. AccMut has much higher relative overheads
compared to SSE and MS, where 79 extra instructions are executed on
average for one original instruction. However, the absolute overheads
of AccMut is even lower, as more redundant computations are removed.
Please note that, despite the overhead, all three approaches are
much faster than plain mutation analysis without any acceleration because of the cost from
compiling. \tblref{tab:plain} shows the execution time for mutation
schemata and plain mutation analysis for the 100 tests of the two
subjects. As we can see, mutation schemata is on average 9.62 times faster.
}

\begin{table}[t]
	\centering
	\caption{Execution Time of Mutation Schemata and Plain
          Mutation Analysis}
	\label{tab:plain}
	\begin{tabular}{|c|c|c|c|}
		\hline
		Subjects & MS & Plain & Plain/MS \\
		\hline
                tcas & 50s & 323s & 6.46X \\
          printtoken & 36s & 504s & 14.00X \\
          \hline
		Total  & 86s & 827s & 9.62X \\
		\hline
	\end{tabular}\\
\end{table}


\subsection{Threats to Validity}

The main threat to internal validity is that our implementations may be wrong. To reduce this threat, we manually checked part of the analysis result and found that the result of our approach is consistent with the result of plain mutation analysis. 

The main threat to external validity is the mutation operators we used
in our experiments. 
Using different mutation operators may have a noticeable effect on the
performance\revise{ of AccMut. For example, a mutation operator designed for
Boolean expressions may have a high chance to produce mutated
statement that is equivalent to the original modulo a state than a
mutation operation on integer expressions. Nevertheless, our
experiment used mutation operators from widely-used tools, and chose
subjects from different areas. Thus, our results have a high chance to
represent the typical use cases.}{. AccMut used mutation operators from widely-used tools, and chose
subjects from different areas. Thus, our results have a high chance to
represent the typical use cases.}

The main threat to construct validity is the way we measure
performance may be imprecise. To reduce this threat, we performed only
sequential but not parallel operations in the experiments, and repeated
the experiments three times and report the median results. As a matter
of fact, there were only very small differences between the three
executions in our experiments, indicating that sequential execution
leads to a stable result.



	\section{Limitations and Future Work}
	\subsection{Other IO Operations} One limitation of the current implementation is that it cannot well
	handle all types of external resources, such as database connections,
	network commutations, etc. To deal with this problem, we need to
	implement the copy-on-write mechanism also for more types of external
	resources. This is a future work on the tool implementation.
	Nevertheless, this limitation may not be a serious one as
	well-written tests often use mock objects rather than directly
	accessing external resources.

        \subsection{Multi-Threaded Programs} So far we only consider
        single-threaded programs. The problem of multi-threaded
        programs is that we need to ensure all threads are properly
        forked when we perform the \code{fork} operation. POSIX \code{fork}
        does not support multi-threaded programs and new mechanisms
        need to be found. This remains a
        future work to be explored.
	
	\subsection{Further Removing Redundancies} Though AccMut removes more
	redundancies in execution than previous approaches, it cannot remove
	all redundancies. A typical case is that two processes may first take
	two different states, and later their execution states become the
	same. To deeply reduce such redundancies, people have resorted to
	sophisticated techniques such as model
	checking~\cite{meinicke2014varexj,KvEPARO:FOSD12} or specialized
	interpreters~\cite{kim2012shared,KvEPARO:FOSD12}. However, such sophisticated
	techniques often impose a large amount of overheads, and it remains as
	future work to explore how to balance between the overheads and the
	benefits.

	\subsection{More Applications} In this paper, we have
        described an implementation of AccMut on first-order mutation
        analysis. The algorithm of AccMut is not limited to
        first-order mutation analysis and can be applied to more
        applications such as high-order mutation analysis, program
        repair, and/or software product line testing. The key is to
        find an efficient way to implement the operations such as
        \filterv and \filterm. These directions also remain as future
        work.

\section{Conclusion}
In this paper, we propose AccMut, which removes redundancies in
executions by exploiting the equivalence of statements modulo the current state.  
The experimental results suggest that our approach can achieve
significant speedup over existing reduction approaches.
The results suggest that there is still a lot of space in reducing redundancies in
mutation analysis, and call for future work in this area. 

\wdel{
\section{Acknowledgements}
	We acknowledge the fruitful
	discussion with Rene Just at University of Massachusetts Amherst and Micheal Ernst at
	University of Washington. This work is supported by the National Key
	Research and Development Program under Grant No. 2016YFB1000105, and
	the National Natural Science Foundation of China under Grant No.
	61421091, 61529201, 61672045, 61332010.
}

\bibliographystyle{ACM-Reference-Format}
\bibliography{reference} 


\begin{thebibliography}{00}


\ifx \showCODEN    \undefined \def \showCODEN     #1{\unskip}     \fi
\ifx \showDOI      \undefined \def \showDOI       #1{{\tt DOI:}\penalty0{#1}\ }
  \fi
\ifx \showISBNx    \undefined \def \showISBNx     #1{\unskip}     \fi
\ifx \showISBNxiii \undefined \def \showISBNxiii  #1{\unskip}     \fi
\ifx \showISSN     \undefined \def \showISSN      #1{\unskip}     \fi
\ifx \showLCCN     \undefined \def \showLCCN      #1{\unskip}     \fi
\ifx \shownote     \undefined \def \shownote      #1{#1}          \fi
\ifx \showarticletitle \undefined \def \showarticletitle #1{#1}   \fi
\ifx \showURL      \undefined \def \showURL       #1{#1}          \fi
\providecommand\bibfield[2]{#2}
\providecommand\bibinfo[2]{#2}
\providecommand\natexlab[1]{#1}
\providecommand\showeprint[2][]{arXiv:#2}

\bibitem[\protect\citeauthoryear{Andrews, Briand, and Labiche}{Andrews
  et~al\mbox{.}}{2005}]%
        {Andrews:05}
\bibfield{author}{\bibinfo{person}{J.~H. Andrews}, \bibinfo{person}{L.~C.
  Briand}, {and} \bibinfo{person}{Y. Labiche}.}
  \bibinfo{year}{2005}\natexlab{}.
\newblock \showarticletitle{Is mutation an appropriate tool for testing
  experiments?}. In \bibinfo{booktitle}{{\em Proc. ICSE}}.
  \bibinfo{pages}{402--411}.
\newblock


\bibitem[\protect\citeauthoryear{Bryant, David~Richard, and
  David~Richard}{Bryant et~al\mbox{.}}{2003}]%
        {architecturebook}
\bibfield{author}{\bibinfo{person}{Randal~E Bryant},
  \bibinfo{person}{O'Hallaron David~Richard}, {and} \bibinfo{person}{O'Hallaron
  David~Richard}.} \bibinfo{year}{2003}\natexlab{}.
\newblock \bibinfo{booktitle}{{\em Computer systems: a programmer's
  perspective}}.
\newblock \bibinfo{publisher}{Prentice Hall Upper Saddle River}.
\newblock


\bibitem[\protect\citeauthoryear{Chekam, Papadakis, and Traon}{Chekam
  et~al\mbox{.}}{2016}]%
        {chekam2016assessing}
\bibfield{author}{\bibinfo{person}{Thierry~Titcheu Chekam},
  \bibinfo{person}{Mike Papadakis}, {and} \bibinfo{person}{Yves~Le Traon}.}
  \bibinfo{year}{2016}\natexlab{}.
\newblock \showarticletitle{Assessing and Comparing Mutation-based Fault
  Localization Techniques}.
\newblock \bibinfo{journal}{{\em arXiv preprint arXiv:1607.05512\/}}
  (\bibinfo{year}{2016}).
\newblock


\bibitem[\protect\citeauthoryear{DeMillo, Lipton, and Sayward}{DeMillo
  et~al\mbox{.}}{1978}]%
        {demillo1978hints}
\bibfield{author}{\bibinfo{person}{Richard~A DeMillo},
  \bibinfo{person}{Richard~J Lipton}, {and} \bibinfo{person}{Frederick~G
  Sayward}.} \bibinfo{year}{1978}\natexlab{}.
\newblock \showarticletitle{Hints on test data selection: Help for the
  practicing programmer}.
\newblock \bibinfo{journal}{{\em Computer\/}} \bibinfo{volume}{11},
  \bibinfo{number}{4} (\bibinfo{year}{1978}), \bibinfo{pages}{34--41}.
\newblock


\bibitem[\protect\citeauthoryear{Devroey, Perrouin, Cordy, Papadakis, Legay,
  and Schobbens}{Devroey et~al\mbox{.}}{2014}]%
        {Devroey:2014:VPM:2635868.2666610}
\bibfield{author}{\bibinfo{person}{Xavier Devroey}, \bibinfo{person}{Gilles
  Perrouin}, \bibinfo{person}{Maxime Cordy}, \bibinfo{person}{Mike Papadakis},
  \bibinfo{person}{Axel Legay}, {and} \bibinfo{person}{Pierre-Yves Schobbens}.}
  \bibinfo{year}{2014}\natexlab{}.
\newblock \showarticletitle{A Variability Perspective of Mutation Analysis}. In
  \bibinfo{booktitle}{{\em FSE}}. \bibinfo{pages}{841--844}.
\newblock


\bibitem[\protect\citeauthoryear{Devroey, Perrouin, Papadakis, Legay,
  Schobbens, and Heymans}{Devroey et~al\mbox{.}}{2016}]%
        {devroey2016featured}
\bibfield{author}{\bibinfo{person}{Xavier Devroey}, \bibinfo{person}{Gilles
  Perrouin}, \bibinfo{person}{Mike Papadakis}, \bibinfo{person}{Axel Legay},
  \bibinfo{person}{Pierre-Yves Schobbens}, {and} \bibinfo{person}{Patrick
  Heymans}.} \bibinfo{year}{2016}\natexlab{}.
\newblock \showarticletitle{Featured model-based mutation analysis}. In
  \bibinfo{booktitle}{{\em ICSE}}. \bibinfo{pages}{655--666}.
\newblock


\bibitem[\protect\citeauthoryear{Do, Elbaum, and Rothermel}{Do
  et~al\mbox{.}}{2005}]%
        {sir}
\bibfield{author}{\bibinfo{person}{Hyunsook Do}, \bibinfo{person}{Sebastian~G.
  Elbaum}, {and} \bibinfo{person}{Gregg Rothermel}.}
  \bibinfo{year}{2005}\natexlab{}.
\newblock \showarticletitle{Supporting Controlled Experimentation with Testing
  Techniques: An Infrastructure and its Potential Impact.}
\newblock \bibinfo{journal}{{\em Empirical Software Engineering: An
  International Journal\/}} \bibinfo{volume}{10}, \bibinfo{number}{4}
  (\bibinfo{year}{2005}), \bibinfo{pages}{405--435}.
\newblock


\bibitem[\protect\citeauthoryear{Durelli, Offutt, and Delamaro}{Durelli
  et~al\mbox{.}}{2012}]%
        {durelli2012toward}
\bibfield{author}{\bibinfo{person}{Vinicius~HS Durelli}, \bibinfo{person}{Jeff
  Offutt}, {and} \bibinfo{person}{Marcio~E Delamaro}.}
  \bibinfo{year}{2012}\natexlab{}.
\newblock \showarticletitle{Toward harnessing high-level language virtual
  machines for further speeding up weak mutation testing}. In
  \bibinfo{booktitle}{{\em ICST}}. \bibinfo{pages}{681--690}.
\newblock


\bibitem[\protect\citeauthoryear{Fraser and Arcuri}{Fraser and Arcuri}{2015}]%
        {fraser2015achieving}
\bibfield{author}{\bibinfo{person}{Gordon Fraser} {and} \bibinfo{person}{Andrea
  Arcuri}.} \bibinfo{year}{2015}\natexlab{}.
\newblock \showarticletitle{Achieving scalable mutation-based generation of
  whole test suites}.
\newblock \bibinfo{journal}{{\em Empirical Software Engineering\/}}
  \bibinfo{volume}{20}, \bibinfo{number}{3} (\bibinfo{year}{2015}),
  \bibinfo{pages}{783--812}.
\newblock


\bibitem[\protect\citeauthoryear{Fraser and Zeller}{Fraser and Zeller}{2012}]%
        {fraser2012mutation}
\bibfield{author}{\bibinfo{person}{Gordon Fraser} {and}
  \bibinfo{person}{Andreas Zeller}.} \bibinfo{year}{2012}\natexlab{}.
\newblock \showarticletitle{Mutation-driven generation of unit tests and
  oracles}.
\newblock \bibinfo{journal}{{\em IEEE Transactions on Software Engineering\/}}
  \bibinfo{volume}{38}, \bibinfo{number}{2} (\bibinfo{year}{2012}),
  \bibinfo{pages}{278--292}.
\newblock


\bibitem[\protect\citeauthoryear{Galeotti, Furia, May, Fraser, and
  Zeller}{Galeotti et~al\mbox{.}}{2015}]%
        {galeotti2015inferring}
\bibfield{author}{\bibinfo{person}{Juan~P Galeotti}, \bibinfo{person}{Carlo~A
  Furia}, \bibinfo{person}{Eva May}, \bibinfo{person}{Gordon Fraser}, {and}
  \bibinfo{person}{Andreas Zeller}.} \bibinfo{year}{2015}\natexlab{}.
\newblock \showarticletitle{Inferring loop invariants by mutation, dynamic
  analysis, and static checking}.
\newblock \bibinfo{journal}{{\em IEEE Transactions on Software Engineering\/}}
  \bibinfo{volume}{41}, \bibinfo{number}{10} (\bibinfo{year}{2015}),
  \bibinfo{pages}{1019--1037}.
\newblock


\bibitem[\protect\citeauthoryear{Gligoric, Zhang, Pereira, and Pokam}{Gligoric
  et~al\mbox{.}}{2013}]%
        {gligoric2013selective}
\bibfield{author}{\bibinfo{person}{Milos Gligoric}, \bibinfo{person}{Lingming
  Zhang}, \bibinfo{person}{Cristiano Pereira}, {and} \bibinfo{person}{Gilles
  Pokam}.} \bibinfo{year}{2013}\natexlab{}.
\newblock \showarticletitle{Selective mutation testing for concurrent code}. In
  \bibinfo{booktitle}{{\em Proc. ISSTA}}. \bibinfo{pages}{224--234}.
\newblock


\bibitem[\protect\citeauthoryear{Gopinath, Alipour, Ahmed, Jensen, and
  Groce}{Gopinath et~al\mbox{.}}{2016a}]%
        {gopinath2016measuring}
\bibfield{author}{\bibinfo{person}{Rahul Gopinath}, \bibinfo{person}{Amin
  Alipour}, \bibinfo{person}{Iftekhar Ahmed}, \bibinfo{person}{Carlos Jensen},
  {and} \bibinfo{person}{Alex Groce}.} \bibinfo{year}{2016}\natexlab{a}.
\newblock \showarticletitle{Measuring Effectiveness of Mutant Sets}. In
  \bibinfo{booktitle}{{\em IEEE ICSTW}}. \bibinfo{pages}{132--141}.
\newblock


\bibitem[\protect\citeauthoryear{Gopinath, Alipour, Ahmed, Jensen, and
  Groce}{Gopinath et~al\mbox{.}}{2016b}]%
        {gopinath2016limits}
\bibfield{author}{\bibinfo{person}{Rahul Gopinath},
  \bibinfo{person}{Mohammad~Amin Alipour}, \bibinfo{person}{Iftekhar Ahmed},
  \bibinfo{person}{Carlos Jensen}, {and} \bibinfo{person}{Alex Groce}.}
  \bibinfo{year}{2016}\natexlab{b}.
\newblock \showarticletitle{On the limits of mutation reduction strategies}. In
  \bibinfo{booktitle}{{\em ICSE}}. \bibinfo{pages}{511--522}.
\newblock


\bibitem[\protect\citeauthoryear{Gopinath, Jensen, and Groce}{Gopinath
  et~al\mbox{.}}{2016c}]%
        {gopinath2016topsy}
\bibfield{author}{\bibinfo{person}{Rahul Gopinath}, \bibinfo{person}{Carlos
  Jensen}, {and} \bibinfo{person}{Alex Groce}.}
  \bibinfo{year}{2016}\natexlab{c}.
\newblock \showarticletitle{Topsy-{Turvy}: a smarter and faster parallelization
  of mutation analysis}. In \bibinfo{booktitle}{{\em ICSE}}.
  \bibinfo{pages}{740--743}.
\newblock


\bibitem[\protect\citeauthoryear{Hamlet}{Hamlet}{1977}]%
        {hamlet1977testing}
\bibfield{author}{\bibinfo{person}{Richard~G Hamlet}.}
  \bibinfo{year}{1977}\natexlab{}.
\newblock \showarticletitle{Testing programs with the aid of a compiler}.
\newblock \bibinfo{journal}{{\em Software Engineering, IEEE Transactions on\/}}
  \bibinfo{volume}{SE-3}, \bibinfo{number}{4} (\bibinfo{year}{1977}),
  \bibinfo{pages}{279--290}.
\newblock


\bibitem[\protect\citeauthoryear{Harman, Jia, Reales~Mateo, and Polo}{Harman
  et~al\mbox{.}}{2014}]%
        {harman2014angels}
\bibfield{author}{\bibinfo{person}{Mark Harman}, \bibinfo{person}{Yue Jia},
  \bibinfo{person}{Pedro Reales~Mateo}, {and} \bibinfo{person}{Macario Polo}.}
  \bibinfo{year}{2014}\natexlab{}.
\newblock \showarticletitle{Angels and monsters: An empirical investigation of
  potential test effectiveness and efficiency improvement from strongly
  subsuming higher order mutation}. In \bibinfo{booktitle}{{\em ASE}}.
  \bibinfo{pages}{397--408}.
\newblock


\bibitem[\protect\citeauthoryear{Hong, Lee, Kwak, Jeon, Ko, Kim, and Kim}{Hong
  et~al\mbox{.}}{2015}]%
        {hong2015mutation}
\bibfield{author}{\bibinfo{person}{Shin Hong}, \bibinfo{person}{Byeongcheol
  Lee}, \bibinfo{person}{Taehoon Kwak}, \bibinfo{person}{Yiru Jeon},
  \bibinfo{person}{Bongsuk Ko}, \bibinfo{person}{Yunho Kim}, {and}
  \bibinfo{person}{Moonzoo Kim}.} \bibinfo{year}{2015}\natexlab{}.
\newblock \showarticletitle{Mutation-based fault localization for real-world
  multilingual programs (T)}. In \bibinfo{booktitle}{{\em ASE}}.
  \bibinfo{pages}{464--475}.
\newblock


\bibitem[\protect\citeauthoryear{Howden}{Howden}{1982}]%
        {howden1982weak}
\bibfield{author}{\bibinfo{person}{William~E. Howden}.}
  \bibinfo{year}{1982}\natexlab{}.
\newblock \showarticletitle{Weak mutation testing and completeness of test
  sets}.
\newblock \bibinfo{journal}{{\em IEEE Transactions on Software Engineering\/}}
  \bibinfo{volume}{SE-8}, \bibinfo{number}{4} (\bibinfo{year}{1982}),
  \bibinfo{pages}{371--379}.
\newblock


\bibitem[\protect\citeauthoryear{Ji, Chen, Xu, and Zhao}{Ji
  et~al\mbox{.}}{2009}]%
        {ji2009novel}
\bibfield{author}{\bibinfo{person}{Changbin Ji}, \bibinfo{person}{Zhenyu Chen},
  \bibinfo{person}{Baowen Xu}, {and} \bibinfo{person}{Zhihong Zhao}.}
  \bibinfo{year}{2009}\natexlab{}.
\newblock \showarticletitle{A Novel Method of Mutation Clustering Based on
  Domain Analysis.}. In \bibinfo{booktitle}{{\em SEKE}}.
  \bibinfo{pages}{422--425}.
\newblock


\bibitem[\protect\citeauthoryear{Jia and Harman}{Jia and Harman}{2011}]%
        {jia2011analysis}
\bibfield{author}{\bibinfo{person}{Yue Jia} {and} \bibinfo{person}{Mark
  Harman}.} \bibinfo{year}{2011}\natexlab{}.
\newblock \showarticletitle{An analysis and survey of the development of
  mutation testing}.
\newblock \bibinfo{journal}{{\em Software Engineering, IEEE Transactions on\/}}
  \bibinfo{volume}{37}, \bibinfo{number}{5} (\bibinfo{year}{2011}),
  \bibinfo{pages}{649--678}.
\newblock


\bibitem[\protect\citeauthoryear{Just}{Just}{2014}]%
        {Just:2014:MMF:2610384.2628053}
\bibfield{author}{\bibinfo{person}{Ren{\'e} Just}.}
  \bibinfo{year}{2014}\natexlab{}.
\newblock \showarticletitle{The {Major} Mutation Framework: Efficient and
  Scalable Mutation Analysis for {Java}}. In \bibinfo{booktitle}{{\em ISSTA}}.
  \bibinfo{publisher}{ACM}, \bibinfo{pages}{433--436}.
\newblock
\showISBNx{978-1-4503-2645-2}


\bibitem[\protect\citeauthoryear{Just, Ernst, and Fraser}{Just
  et~al\mbox{.}}{2014a}]%
        {just2014efficient}
\bibfield{author}{\bibinfo{person}{Ren{\'e} Just}, \bibinfo{person}{Michael~D
  Ernst}, {and} \bibinfo{person}{Gordon Fraser}.}
  \bibinfo{year}{2014}\natexlab{a}.
\newblock \showarticletitle{Efficient mutation analysis by propagating and
  partitioning infected execution states}. In \bibinfo{booktitle}{{\em ISSTA}}.
  \bibinfo{pages}{315--326}.
\newblock


\bibitem[\protect\citeauthoryear{Just, Jalali, Inozemtseva, Ernst, Holmes, and
  Fraser}{Just et~al\mbox{.}}{2014b}]%
        {just2014mutants}
\bibfield{author}{\bibinfo{person}{Ren{\'e} Just}, \bibinfo{person}{Darioush
  Jalali}, \bibinfo{person}{Laura Inozemtseva}, \bibinfo{person}{Michael~D
  Ernst}, \bibinfo{person}{Reid Holmes}, {and} \bibinfo{person}{Gordon
  Fraser}.} \bibinfo{year}{2014}\natexlab{b}.
\newblock \showarticletitle{Are mutants a valid substitute for real faults in
  software testing?}. In \bibinfo{booktitle}{{\em FSE}}.
  \bibinfo{pages}{654--665}.
\newblock


\bibitem[\protect\citeauthoryear{Just, Kapfhammer, and Schweiggert}{Just
  et~al\mbox{.}}{2012}]%
        {just2012using}
\bibfield{author}{\bibinfo{person}{Ren{\'e} Just}, \bibinfo{person}{Gregory~M
  Kapfhammer}, {and} \bibinfo{person}{Franz Schweiggert}.}
  \bibinfo{year}{2012}\natexlab{}.
\newblock \showarticletitle{Using non-redundant mutation operators and test
  suite prioritization to achieve efficient and scalable mutation analysis}. In
  \bibinfo{booktitle}{{\em ISSRE}}. \bibinfo{pages}{11--20}.
\newblock


\bibitem[\protect\citeauthoryear{Just, Schweiggert, and Kapfhammer}{Just
  et~al\mbox{.}}{2011}]%
        {just2011major}
\bibfield{author}{\bibinfo{person}{Rene Just}, \bibinfo{person}{Franz
  Schweiggert}, {and} \bibinfo{person}{Gregory~M Kapfhammer}.}
  \bibinfo{year}{2011}\natexlab{}.
\newblock \showarticletitle{MAJOR: An efficient and extensible tool for
  mutation analysis in a {Java} compiler}. In \bibinfo{booktitle}{{\em ASE}}.
  \bibinfo{pages}{612--615}.
\newblock


\bibitem[\protect\citeauthoryear{K{\"a}stner, von Rhein, Erdweg, Pusch, Apel,
  Rendel, and Ostermann}{K{\"a}stner et~al\mbox{.}}{2012}]%
        {KvEPARO:FOSD12}
\bibfield{author}{\bibinfo{person}{Christian K{\"a}stner},
  \bibinfo{person}{Alexander von Rhein}, \bibinfo{person}{Sebastian Erdweg},
  \bibinfo{person}{Jonas Pusch}, \bibinfo{person}{Sven Apel},
  \bibinfo{person}{Tillmann Rendel}, {and} \bibinfo{person}{Klaus Ostermann}.}
  \bibinfo{year}{2012}\natexlab{}.
\newblock \showarticletitle{Toward Variability-Aware Testing}. In
  \bibinfo{booktitle}{{\em FOSD}}. \bibinfo{pages}{1--8}.
\newblock


\bibitem[\protect\citeauthoryear{Kim, Khurshid, and Batory}{Kim
  et~al\mbox{.}}{2012}]%
        {kim2012shared}
\bibfield{author}{\bibinfo{person}{Chang Hwan~Peter Kim},
  \bibinfo{person}{Sarfraz Khurshid}, {and} \bibinfo{person}{Don Batory}.}
  \bibinfo{year}{2012}\natexlab{}.
\newblock \showarticletitle{Shared execution for efficiently testing product
  lines}. In \bibinfo{booktitle}{{\em ISSRE}}. \bibinfo{pages}{221--230}.
\newblock


\bibitem[\protect\citeauthoryear{Kim, Nam, Song, and Kim}{Kim
  et~al\mbox{.}}{2013}]%
        {PAR}
\bibfield{author}{\bibinfo{person}{Dongsun Kim}, \bibinfo{person}{Jaechang
  Nam}, \bibinfo{person}{Jaewoo Song}, {and} \bibinfo{person}{Sunghun Kim}.}
  \bibinfo{year}{2013}\natexlab{}.
\newblock \showarticletitle{Automatic patch generation learned from
  human-written patches}. In \bibinfo{booktitle}{{\em ICSE '13}}.
  \bibinfo{pages}{802--811}.
\newblock


\bibitem[\protect\citeauthoryear{King and Offutt}{King and Offutt}{1991}]%
        {King:1991:FLS:116633.116640}
\bibfield{author}{\bibinfo{person}{K.~N. King} {and}
  \bibinfo{person}{A.~Jefferson Offutt}.} \bibinfo{year}{1991}\natexlab{}.
\newblock \showarticletitle{A Fortran Language System for Mutation-based
  Software Testing}.
\newblock \bibinfo{journal}{{\em Softw. Pract. Exper.\/}} \bibinfo{volume}{21},
  \bibinfo{number}{7} (\bibinfo{date}{June} \bibinfo{year}{1991}),
  \bibinfo{pages}{685--718}.
\newblock
\showISSN{0038-0644}


\bibitem[\protect\citeauthoryear{Krauser, Mathur, and Rego}{Krauser
  et~al\mbox{.}}{1991}]%
        {krauser1991high}
\bibfield{author}{\bibinfo{person}{Edward~W. Krauser},
  \bibinfo{person}{Aditya~P. Mathur}, {and} \bibinfo{person}{Vernon~J Rego}.}
  \bibinfo{year}{1991}\natexlab{}.
\newblock \showarticletitle{High performance software testing on SIMD
  machines}.
\newblock \bibinfo{journal}{{\em IEEE Transactions on Software Engineering\/}}
  \bibinfo{volume}{17}, \bibinfo{number}{5} (\bibinfo{year}{1991}),
  \bibinfo{pages}{403--423}.
\newblock


\bibitem[\protect\citeauthoryear{Kurtz, Ammann, Offutt, Delamaro, Kurtz, and
  G{\"o}k{\c{c}}e}{Kurtz et~al\mbox{.}}{2016}]%
        {kurtz2016analyzing}
\bibfield{author}{\bibinfo{person}{Bob Kurtz}, \bibinfo{person}{Paul Ammann},
  \bibinfo{person}{Jeff Offutt}, \bibinfo{person}{M{\'a}rcio~E Delamaro},
  \bibinfo{person}{Mariet Kurtz}, {and} \bibinfo{person}{Nida
  G{\"o}k{\c{c}}e}.} \bibinfo{year}{2016}\natexlab{}.
\newblock \showarticletitle{Analyzing the validity of selective mutation with
  dominator mutants}. In \bibinfo{booktitle}{{\em FSE}}.
  \bibinfo{pages}{571--582}.
\newblock


\bibitem[\protect\citeauthoryear{Lattner and Adve}{Lattner and Adve}{2004}]%
        {lattner2004llvm}
\bibfield{author}{\bibinfo{person}{Chris Lattner} {and} \bibinfo{person}{Vikram
  Adve}.} \bibinfo{year}{2004}\natexlab{}.
\newblock \showarticletitle{{LLVM}: A compilation framework for lifelong
  program analysis \& transformation}. In \bibinfo{booktitle}{{\em CGO}}.
  \bibinfo{pages}{75--86}.
\newblock


\bibitem[\protect\citeauthoryear{Martin and Xie}{Martin and Xie}{2007}]%
        {martin2007fault}
\bibfield{author}{\bibinfo{person}{Evan Martin} {and} \bibinfo{person}{Tao
  Xie}.} \bibinfo{year}{2007}\natexlab{}.
\newblock \showarticletitle{A fault model and mutation testing of access
  control policies}. In \bibinfo{booktitle}{{\em WWW}}.
  \bibinfo{pages}{667--676}.
\newblock


\bibitem[\protect\citeauthoryear{Meinicke}{Meinicke}{2014}]%
        {meinicke2014varexj}
\bibfield{author}{\bibinfo{person}{Jens Meinicke}.}
  \bibinfo{year}{2014}\natexlab{}.
\newblock \bibinfo{title}{VarexJ: A Variability-Aware Interpreter for {Java}
  Applications}.
\newblock   (\bibinfo{year}{2014}).
\newblock


\bibitem[\protect\citeauthoryear{Mirshokraie, Mesbah, and
  Pattabiraman}{Mirshokraie et~al\mbox{.}}{2015}]%
        {mirshokraie2015guided}
\bibfield{author}{\bibinfo{person}{Shabnam Mirshokraie}, \bibinfo{person}{Ali
  Mesbah}, {and} \bibinfo{person}{Karthik Pattabiraman}.}
  \bibinfo{year}{2015}\natexlab{}.
\newblock \showarticletitle{Guided mutation testing for javascript web
  applications}.
\newblock \bibinfo{journal}{{\em IEEE Transactions on Software Engineering\/}}
  \bibinfo{volume}{41}, \bibinfo{number}{5} (\bibinfo{year}{2015}),
  \bibinfo{pages}{429--444}.
\newblock


\bibitem[\protect\citeauthoryear{Moon, Kim, Kim, and Yoo}{Moon
  et~al\mbox{.}}{2014}]%
        {moon2014ask}
\bibfield{author}{\bibinfo{person}{Seokhyeon Moon}, \bibinfo{person}{Yunho
  Kim}, \bibinfo{person}{Moonzoo Kim}, {and} \bibinfo{person}{Shin Yoo}.}
  \bibinfo{year}{2014}\natexlab{}.
\newblock \showarticletitle{Ask the mutants: Mutating faulty programs for fault
  localization}. In \bibinfo{booktitle}{{\em ICST}}. \bibinfo{pages}{153--162}.
\newblock


\bibitem[\protect\citeauthoryear{Offutt and Craft}{Offutt and Craft}{1994}]%
        {offutt1994using}
\bibfield{author}{\bibinfo{person}{A~Jefferson Offutt} {and}
  \bibinfo{person}{W~Michael Craft}.} \bibinfo{year}{1994}\natexlab{}.
\newblock \showarticletitle{Using compiler optimization techniques to detect
  equivalent mutants}.
\newblock \bibinfo{journal}{{\em Software Testing, Verification and
  Reliability\/}} \bibinfo{volume}{4}, \bibinfo{number}{3}
  (\bibinfo{year}{1994}), \bibinfo{pages}{131--154}.
\newblock


\bibitem[\protect\citeauthoryear{Offutt, Pargas, Fichter, and Khambekar}{Offutt
  et~al\mbox{.}}{1992}]%
        {offutt1992mutation}
\bibfield{author}{\bibinfo{person}{A~Jefferson Offutt}, \bibinfo{person}{Roy~P
  Pargas}, \bibinfo{person}{Scott~V Fichter}, {and} \bibinfo{person}{Prashant~K
  Khambekar}.} \bibinfo{year}{1992}\natexlab{}.
\newblock \showarticletitle{Mutation testing of software using a MIMD
  computer}. In \bibinfo{booktitle}{{\em Proc. ICPP}}.
\newblock


\bibitem[\protect\citeauthoryear{Papadakis, Henard, Harman, Jia, and
  Le~Traon}{Papadakis et~al\mbox{.}}{2016}]%
        {papadakis2016threats}
\bibfield{author}{\bibinfo{person}{Mike Papadakis},
  \bibinfo{person}{Christopher Henard}, \bibinfo{person}{Mark Harman},
  \bibinfo{person}{Yue Jia}, {and} \bibinfo{person}{Yves Le~Traon}.}
  \bibinfo{year}{2016}\natexlab{}.
\newblock \showarticletitle{Threats to the validity of mutation-based test
  assessment}. In \bibinfo{booktitle}{{\em ISSTA}}. \bibinfo{pages}{354--365}.
\newblock


\bibitem[\protect\citeauthoryear{Papadakis, Henard, and Le~Traon}{Papadakis
  et~al\mbox{.}}{2014}]%
        {papadakis2014sampling}
\bibfield{author}{\bibinfo{person}{Mike Papadakis},
  \bibinfo{person}{Christopher Henard}, {and} \bibinfo{person}{Yves Le~Traon}.}
  \bibinfo{year}{2014}\natexlab{}.
\newblock \showarticletitle{Sampling program inputs with mutation analysis:
  Going beyond combinatorial interaction testing}. In \bibinfo{booktitle}{{\em
  ICST}}. \bibinfo{pages}{1--10}.
\newblock


\bibitem[\protect\citeauthoryear{Papadakis, Jia, Harman, and
  Le~Traon}{Papadakis et~al\mbox{.}}{2015}]%
        {papadakis2015trivial}
\bibfield{author}{\bibinfo{person}{Mike Papadakis}, \bibinfo{person}{Yue Jia},
  \bibinfo{person}{Mark Harman}, {and} \bibinfo{person}{Yves Le~Traon}.}
  \bibinfo{year}{2015}\natexlab{}.
\newblock \showarticletitle{Trivial compiler equivalence: A large scale
  empirical study of a simple, fast and effective equivalent mutant detection
  technique}. In \bibinfo{booktitle}{{\em ICSE}}. \bibinfo{pages}{936--946}.
\newblock


\bibitem[\protect\citeauthoryear{Papadakis and Le~Traon}{Papadakis and
  Le~Traon}{2012}]%
        {papadakis2012using}
\bibfield{author}{\bibinfo{person}{Mike Papadakis} {and} \bibinfo{person}{Yves
  Le~Traon}.} \bibinfo{year}{2012}\natexlab{}.
\newblock \showarticletitle{Using mutants to locate" unknown" faults}. In
  \bibinfo{booktitle}{{\em ICST}}. \bibinfo{pages}{691--700}.
\newblock


\bibitem[\protect\citeauthoryear{Papadakis and Le~Traon}{Papadakis and
  Le~Traon}{2014}]%
        {papadakis2014effective}
\bibfield{author}{\bibinfo{person}{Mike Papadakis} {and} \bibinfo{person}{Yves
  Le~Traon}.} \bibinfo{year}{2014}\natexlab{}.
\newblock \showarticletitle{Effective fault localization via mutation analysis:
  A selective mutation approach}. In \bibinfo{booktitle}{{\em Proceedings of
  the 29th Annual ACM Symposium on Applied Computing}}.
  \bibinfo{pages}{1293--1300}.
\newblock


\bibitem[\protect\citeauthoryear{Papadakis and Le~Traon}{Papadakis and
  Le~Traon}{2015}]%
        {papadakis2015metallaxis}
\bibfield{author}{\bibinfo{person}{Mike Papadakis} {and} \bibinfo{person}{Yves
  Le~Traon}.} \bibinfo{year}{2015}\natexlab{}.
\newblock \showarticletitle{Metallaxis-FL: mutation-based fault localization}.
\newblock \bibinfo{journal}{{\em Software Testing, Verification and
  Reliability\/}} \bibinfo{volume}{25}, \bibinfo{number}{5-7}
  (\bibinfo{year}{2015}), \bibinfo{pages}{605--628}.
\newblock


\bibitem[\protect\citeauthoryear{Pearson, Campos, Just, Fraser, Abreu, Ernst,
  Pang, and Keller}{Pearson et~al\mbox{.}}{2017}]%
        {icse17FaultPrediction}
\bibfield{author}{\bibinfo{person}{Spencer Pearson}, \bibinfo{person}{Jos{\'e}
  Campos}, \bibinfo{person}{Ren{\'e} Just}, \bibinfo{person}{Gordon Fraser},
  \bibinfo{person}{Rui Abreu}, \bibinfo{person}{Michael~D. Ernst},
  \bibinfo{person}{Deric Pang}, {and} \bibinfo{person}{Benjamin Keller}.}
  \bibinfo{year}{2017}\natexlab{}.
\newblock \showarticletitle{Evaluating \& improving fault localization
  techniques}. In \bibinfo{booktitle}{{\em ICSE}}.
\newblock


\bibitem[\protect\citeauthoryear{Qi, Mao, Lei, Dai, and Wang}{Qi
  et~al\mbox{.}}{2014}]%
        {RSRepair}
\bibfield{author}{\bibinfo{person}{Yuhua Qi}, \bibinfo{person}{Xiaoguang Mao},
  \bibinfo{person}{Yan Lei}, \bibinfo{person}{Ziying Dai}, {and}
  \bibinfo{person}{Chengsong Wang}.} \bibinfo{year}{2014}\natexlab{}.
\newblock \showarticletitle{The Strength of Random Search on Automated Program
  Repair}. In \bibinfo{booktitle}{{\em Proceedings of the 36th International
  Conference on Software Engineering}} {\em (\bibinfo{series}{ICSE 2014})}.
  \bibinfo{pages}{254--265}.
\newblock


\bibitem[\protect\citeauthoryear{Schuler and Zeller}{Schuler and
  Zeller}{2009}]%
        {schuler2009javalanche}
\bibfield{author}{\bibinfo{person}{David Schuler} {and}
  \bibinfo{person}{Andreas Zeller}.} \bibinfo{year}{2009}\natexlab{}.
\newblock \showarticletitle{Javalanche: efficient mutation testing for {Java}}.
  In \bibinfo{booktitle}{{\em ESEC/FSE}}. \bibinfo{pages}{297--298}.
\newblock


\bibitem[\protect\citeauthoryear{Souza, Papadakis, Le~Traon, and
  Delamaro}{Souza et~al\mbox{.}}{2016}]%
        {souza2016strong}
\bibfield{author}{\bibinfo{person}{Francisco Carlos~M Souza},
  \bibinfo{person}{Mike Papadakis}, \bibinfo{person}{Yves Le~Traon}, {and}
  \bibinfo{person}{M{\'a}rcio~E Delamaro}.} \bibinfo{year}{2016}\natexlab{}.
\newblock \showarticletitle{Strong mutation-based test data generation using
  hill climbing}. In \bibinfo{booktitle}{{\em IEEE/ACM SBST}}.
  \bibinfo{pages}{45--54}.
\newblock


\bibitem[\protect\citeauthoryear{Tokumoto, Yoshida, Sakamoto, and
  Honiden}{Tokumoto et~al\mbox{.}}{2016}]%
        {tokumoto2016muvm}
\bibfield{author}{\bibinfo{person}{Susumu Tokumoto}, \bibinfo{person}{Hiroaki
  Yoshida}, \bibinfo{person}{Kazunori Sakamoto}, {and}
  \bibinfo{person}{Shinichi Honiden}.} \bibinfo{year}{2016}\natexlab{}.
\newblock \showarticletitle{MuVM: Higher Order Mutation Analysis Virtual
  Machine for C}. In \bibinfo{booktitle}{{\em ICST}}.
  \bibinfo{pages}{320--329}.
\newblock


\bibitem[\protect\citeauthoryear{Untch, Offutt, and Harrold}{Untch
  et~al\mbox{.}}{1993}]%
        {untch1993mutation}
\bibfield{author}{\bibinfo{person}{Roland~H Untch},
  \bibinfo{person}{A~Jefferson Offutt}, {and} \bibinfo{person}{Mary~Jean
  Harrold}.} \bibinfo{year}{1993}\natexlab{}.
\newblock \showarticletitle{Mutation analysis using mutant schemata}. In
  \bibinfo{booktitle}{{\em Proc. ISSTA}}. \bibinfo{pages}{139--148}.
\newblock


\bibitem[\protect\citeauthoryear{Weimer, Fry, and Forrest}{Weimer
  et~al\mbox{.}}{2013}]%
        {AE}
\bibfield{author}{\bibinfo{person}{W. Weimer}, \bibinfo{person}{Z.P. Fry},
  {and} \bibinfo{person}{S. Forrest}.} \bibinfo{year}{2013}\natexlab{}.
\newblock \showarticletitle{Leveraging program equivalence for adaptive program
  repair: Models and first results}. In \bibinfo{booktitle}{{\em ASE}}.
  \bibinfo{pages}{356--366}.
\newblock


\bibitem[\protect\citeauthoryear{Weimer, Nguyen, Le~Goues, and Forrest}{Weimer
  et~al\mbox{.}}{2009}]%
        {GenProg}
\bibfield{author}{\bibinfo{person}{Westley Weimer}, \bibinfo{person}{ThanhVu
  Nguyen}, \bibinfo{person}{Claire Le~Goues}, {and} \bibinfo{person}{Stephanie
  Forrest}.} \bibinfo{year}{2009}\natexlab{}.
\newblock \showarticletitle{Automatically finding patches using genetic
  programming}. In \bibinfo{booktitle}{{\em ICSE '09}}.
  \bibinfo{pages}{364--374}.
\newblock


\bibitem[\protect\citeauthoryear{Wong and Mathur}{Wong and Mathur}{1995}]%
        {wong1995reducing}
\bibfield{author}{\bibinfo{person}{W~Eric Wong} {and} \bibinfo{person}{Aditya~P
  Mathur}.} \bibinfo{year}{1995}\natexlab{}.
\newblock \showarticletitle{Reducing the cost of mutation testing: An empirical
  study}.
\newblock \bibinfo{journal}{{\em Journal of Systems and Software\/}}
  \bibinfo{volume}{31}, \bibinfo{number}{3} (\bibinfo{year}{1995}),
  \bibinfo{pages}{185--196}.
\newblock


\bibitem[\protect\citeauthoryear{Yao, Harman, and Jia}{Yao
  et~al\mbox{.}}{2014}]%
        {yao2014study}
\bibfield{author}{\bibinfo{person}{Xiangjuan Yao}, \bibinfo{person}{Mark
  Harman}, {and} \bibinfo{person}{Yue Jia}.} \bibinfo{year}{2014}\natexlab{}.
\newblock \showarticletitle{A study of equivalent and stubborn mutation
  operators using human analysis of equivalence}. In \bibinfo{booktitle}{{\em
  ICSE}}. \bibinfo{pages}{919--930}.
\newblock


\bibitem[\protect\citeauthoryear{Zhang, Wang, Zhang, Hao, Zang, Cheng, and
  Zhang}{Zhang et~al\mbox{.}}{2016}]%
        {zhang2016predictive}
\bibfield{author}{\bibinfo{person}{Jie Zhang}, \bibinfo{person}{Ziyi Wang},
  \bibinfo{person}{Lingming Zhang}, \bibinfo{person}{Dan Hao},
  \bibinfo{person}{Lei Zang}, \bibinfo{person}{Shiyang Cheng}, {and}
  \bibinfo{person}{Lu Zhang}.} \bibinfo{year}{2016}\natexlab{}.
\newblock \showarticletitle{Predictive mutation testing}. In
  \bibinfo{booktitle}{{\em ISSTA}}. \bibinfo{pages}{342--353}.
\newblock


\bibitem[\protect\citeauthoryear{Zhang, Hou, Hu, Xie, and Mei}{Zhang
  et~al\mbox{.}}{2010}]%
        {zhang2010operator}
\bibfield{author}{\bibinfo{person}{Lu Zhang}, \bibinfo{person}{Shan-Shan Hou},
  \bibinfo{person}{Jun-Jue Hu}, \bibinfo{person}{Tao Xie}, {and}
  \bibinfo{person}{Hong Mei}.} \bibinfo{year}{2010}\natexlab{}.
\newblock \showarticletitle{Is operator-based mutant selection superior to
  random mutant selection?}. In \bibinfo{booktitle}{{\em Proc. ICSE}}.
  \bibinfo{pages}{435--444}.
\newblock


\bibitem[\protect\citeauthoryear{Zhang, Marinov, and Khurshid}{Zhang
  et~al\mbox{.}}{2013}]%
        {zhang2013faster}
\bibfield{author}{\bibinfo{person}{Lingming Zhang}, \bibinfo{person}{Darko
  Marinov}, {and} \bibinfo{person}{Sarfraz Khurshid}.}
  \bibinfo{year}{2013}\natexlab{}.
\newblock \showarticletitle{Faster mutation testing inspired by test
  prioritization and reduction}. In \bibinfo{booktitle}{{\em Proc. ISSTA}}.
  \bibinfo{pages}{235--245}.
\newblock


\bibitem[\protect\citeauthoryear{Zhang, Marinov, Zhang, and Khurshid}{Zhang
  et~al\mbox{.}}{2012}]%
        {zhang2012regression}
\bibfield{author}{\bibinfo{person}{Lingming Zhang}, \bibinfo{person}{Darko
  Marinov}, \bibinfo{person}{Lu Zhang}, {and} \bibinfo{person}{Sarfraz
  Khurshid}.} \bibinfo{year}{2012}\natexlab{}.
\newblock \showarticletitle{Regression mutation testing}. In
  \bibinfo{booktitle}{{\em Proc. ISSTA}}. \bibinfo{pages}{331--341}.
\newblock


\bibitem[\protect\citeauthoryear{Zhang, Zhang, and Khurshid}{Zhang
  et~al\mbox{.}}{2013}]%
        {zhang2013injecting}
\bibfield{author}{\bibinfo{person}{Lingming Zhang}, \bibinfo{person}{Lu Zhang},
  {and} \bibinfo{person}{Sarfraz Khurshid}.} \bibinfo{year}{2013}\natexlab{}.
\newblock \showarticletitle{Injecting mechanical faults to localize developer
  faults for evolving software}. In \bibinfo{booktitle}{{\em Proc. OOPSLA}}.
  \bibinfo{pages}{765--784}.
\newblock


\end{thebibliography}

\end{document}